\documentclass{article}
\usepackage{hyperref}
\usepackage{soul}
\usepackage[utf8]{inputenc}
\usepackage{amsmath}
\usepackage{graphicx}
\usepackage[affil-it]{authblk}
\usepackage{fullpage}
\usepackage{longtable}
\usepackage{color}
\usepackage{arydshln}

\usepackage{multirow}
\newcommand{\euro}{\text{€}}
\usepackage{siunitx}
\usepackage{hyperref}
\usepackage{float}
\usepackage{eucal}
\usepackage{textcomp}
\usepackage{booktabs}
\usepackage{makecell}
\usepackage{ctable}
\usepackage{caption}
\usepackage{nicefrac}

\usepackage{array}  
\usepackage{tabularx}  
\usepackage{xcolor}
\usepackage{multirow}

\newlength\myindention
\DeclareCaptionFormat{myformat}
{\hspace*{\myindention}#1#2#3}
\usepackage[numbers,sort&compress,comma]{natbib}

\title{Transition pathways to electrified chemical production within sector-coupled national energy systems}
\author[a]{Patricia Mayer}
\author[a]{Florian Joseph Baader}
\author[a]{David Yang Shu}
\author[a]{Ludger Leenders} 
\author[b]{Christian Zibunas} 
\author[a]{Stefano Moret} 
\author[a]{André Bardow\thanks{corresponding author: abardow@ethz.ch}}

\affil[a]{Energy \& Process Systems Engineering, Department of Mechanical and Process
Engineering, ETH Zürich, Switzerland}
\affil[b]{Institute of Technical Thermodynamics, RWTH Aachen University, Germany}

\date{\today}

\begin{document}

\maketitle

\abstract{\noindent The chemical industry's transition to net-zero greenhouse gas (GHG) emissions is particularly challenging due to the carbon inherently contained in chemical products, eventually released to the environment. Fossil feedstock-based production can be replaced by electrified chemical production, combining carbon capture and utilization (CCU) with electrolysis-based hydrogen.  However, electrified chemical production requires vast amounts of  clean electricity, leading to competition in our sector-coupled energy systems. In this work, we investigate the pathway of the chemical industry towards electrified production within the context of a sector-coupled national energy system's transition to net-zero emissions. Our results show that the sectors for electricity, low-temperature heat, and mobility transition before the chemical industry due to the required build-up of renewables, and to the higher emissions abatement of heat pumps and battery electric vehicles. To achieve the net-zero target, the energy system relies on clean energy imports to cover 41\% of its electricity needs, largely driven by the high energy requirements of a fully electrified chemical industry. Nonetheless, a partially electrified industry combined with dispatchable production alternatives provides flexibility to the energy system by enabling electrified production when renewable electricity is available. Hence, a partially electrified, diversified chemical industry can support the integration of intermittent renewables, serving as a valuable component in net-zero energy systems.}\\

\section{Introduction}
The chemical industry consumes 14\% of oil and 8\% of natural gas supply globally \cite{InternationalEnergyAgency.2018}. Fossil-based resources used in the chemical industry account for 7\% of global greenhouse gas (GHG) emissions \cite{van2013technology}. Thus, defossilizing the chemical industry is crucial for meeting net-zero greenhouse gas (GHG) emissions targets. However, defossilization of the chemical industry is challenging due to the inherent need for carbon in  chemical products, traditionally supplied by fossil-based hydrocarbons. The need for a material input accounts for 58\% of the chemical industry's fossil demand \cite{InternationalEnergyAgency.2017}. The carbon in the feedstock is usually released to the environment either as direct emissions during chemicals production, as is the case for conventional ammonia \cite{mayer2023blue}, or during the chemicals use and end-of-life phases, as is the case for waste incinerating carbon-containing chemicals such as plastics.

One way to reduce the chemical industry’s reliance on fossil-based feedstocks is through electrified production, where the feedstocks are obtained through carbon capture and utilization (CCU) and electrolytic hydrogen ($\mathrm{H_{2}}$) \cite{schiffer2017electrification,Lopez.2023}. With CCU, $\mathrm{CO_{2}}$ is captured from industrial point sources or directly from the air and is used as a carbon-based feedstock for chemicals, thus enabling a circular chemical industry. Electrolytic $\mathrm{H_{2}}$ is produced through water-splitting into $\mathrm{H_{2}}$ and oxygen ($\mathrm{O_{2}}$) by applying electricity, thus avoiding the $\mathrm{CO_{2}}$ emissions from conventional fossil-based $\mathrm{H_{2}}$ production. Using renewable electricity for both CCU and electrolytic $\mathrm{H_{2}}$ can result in a low-emission chemical industry, with a global emissions reduction potential of 3.5 Gt $\mathrm{CO_{2}}$-eq if abundant clean electricity is available \cite{Katelhon.2019}.

CCU-based methanol has been identified as a promising precursor for all high value chemicals \cite{Yarulina.2018} and was shown to safely operate within the Earth's carbon emissions planetary boundaries \cite{GonzalezGaray.2019}. A future highly electrified chemical industry has been predicted to have lower annualized costs than a biomass-based industry, indicating that an electrified chemical industry may be the most economical solution for a future green chemical industry \cite{Lopez.2023}.

Thus, electrification provides a pathway to a sustainable chemical industry, but requires an abundance of electricity that is low in GHG-intensity. Reducing global GHG emissions by 3.5 Gt $\mathrm{CO_{2}}$-eq would require 18.1~PWh of low carbon electricity \cite{Katelhon.2019}, corresponding to 126\% of the global renewable electricity production targets in 2030 even under optimistic assumptions on technology development \cite{Katelhon.2019}. Obtaining a mostly electrified chemical industry by 2050, considering future demand growth, would require nearly 40~PWh of electricity, comprising 150\% of today's global electricity generation \cite{Lopez.2023}. This high demand for clean electricity is the main barrier in the transition to an electrified chemical industry  \cite{MacDowell.2017, InternationalEnergyAgency.2019, bazzanella2017low, Gabrielli.2020,mallapragada2023decarbonization,brethauer2021towards,palm2016electricity}.

Renewables can potentially be expanded massively to supply an electrified chemical industry with sufficient clean electricity by 2050 \cite{IEA.WEO.2023}. However, during the energy transition, clean electricity will be limited and used more efficiently in other sectors. Particularly, single-technology comparisons indicate that heat pumps and battery electric vehicles might reduce emissions more than utilizing the electricity in an electrified chemical industry, which has a low energy return on investment \cite{MacDowell.2017, sternberg2015power}. Thus, the transition towards electrified chemical production must be considered in the context of a sector-coupled energy system.

Considering the combined transition also enables identification of the effects of intermittent renewable electricity supply on electrified chemical production \cite{ganzer2020comparative}, as well as the potential flexibility provision from electrified chemical processes to the energy system \cite{mayer2024flexibility}.  For instance, Almajed et al. \cite{Hodge2024syngas} show that the profitability of electrified syngas production depends on renewable electricity availability and price. With respect to flexibility to the energy system, electrified chemical processes can help manage grid congestion in a renewables-dominated power grid through flexible operation \cite{foslie2024faster}. Chemicals can also provide flexibility over varying timescales, from hours to seasons \cite{mallapragada2023decarbonization}, and have been deemed essential, together with other CCU options, as flexibility providers to renewables-dominated energy systems \cite{mikulvcic2019flexible}. However, as highlighted by Guerra et al. \cite{guerra2023barriers}, the value of this flexibility to the energy system is not well understood, and requires further investigation.
 
Studies have investigated transition pathways of sector-coupled energy systems including industry, showing the importance of resolving the fully coupled system for the costs and technological transitions of each sector \cite{Bogdanov.2021, aboumahboub2020decarbonization, burandt2019decarbonizing, kawai2022role, manuel2022high, martinez2022modelling, pickering2022diversity, ram2022accelerating}. For instance, Bogdanov et al. \cite{Bogdanov.2021} find that including more electrified industry increases flexibility while reducing the system's levelized cost of energy. While insightful regarding important interactions between industry and the energy system, these studies either neglect the chemical industry altogether \cite{aboumahboub2020decarbonization, burandt2019decarbonizing}, or simplify its representation. Simplifications include aggregating individual chemicals into a single chemical product (i.e. aggregated aromatics or high-value chemicals) \cite{kawai2022role, manuel2022high, martinez2022modelling, pickering2022diversity}, or focusing on the electrified feedstock transition rather than chemical products and thus limiting the production options for chemicals further downstream \cite{ram2022accelerating, Bogdanov.2021}. As their focus is not on the chemical industry, these analyses disregard the timing and evolution of the chemical industry's transition, as well as its interactions with the energy system. 

Other works have focused on the interactions between a sustainable chemical industry and other energy sectors \cite{Rixhon.2022, Ioannou.2023,stegmann2022plastic,daioglou2015competing,tsiropoulos2018emerging,hofmann2024h}. However, these studies either neglect the transition pathway and only consider a future chemical industry \cite{Rixhon.2022,Ioannou.2023}, or focus on alternative production routes such as biomass and recycling \cite{stegmann2022plastic,daioglou2015competing, tsiropoulos2018emerging}. While biomass and recycling are promising solutions for GHG mitigation of the chemical industry \cite{zheng2019strategies, meys2021achieving,bachmann2023towards}, they both have limitations such as biomass availability and competition \cite{daioglou2015competing} and recycling ramp-up challenges \cite{ellenMacArthur}. Hence, electrified chemical production is an important component in the portfolio of potential solutions \cite{stegmann2022plastic}.

 In this work, we investigate transition pathways towards an electrified chemical industry while considering the interactions with a transitioning sector-coupled national energy system. We consider the German energy system for our case study due to the importance of its chemical industry, being the third largest industry in Germany \cite{cefic} and the third largest chemical exporter worldwide \cite{statista}. We consider the electricity, residential heat, industrial heat, and private mobility sectors within the energy system \cite{Baumgartner.2021}. For the chemical industry, we consider the seven base and high value chemicals: ammonia, methanol, ethylene, propylene, benzene, xylene, and toluene. More than 90\% of the oil and gas entering the chemical industry as feedstock, by mass, is used for the production these chemicals \cite{InternationalEnergyAgency.2018}. Moreover, the energy requirements for these seven chemicals account for two thirds of the chemical sector's energy consumption \cite{IEAPrimaryChemicalProduction}, making these chemicals a good subset for representing the industry. We gather process data for the electrified chemical production processes from a comprehensive literature review, combining published data with private databases.
 
 We determine the optimal timing of the chemical industry’s transition relative to the transition of the other energy sectors, finding that the chemical industry transitions after the build-up of renewable electricity and the transition to heat pumps and battery electric vehicles. We also take a deep dive into the chemical industry's transition, determining the optimal transition order of the individual chemicals. We introduce the \textit{Cost-Avoided}, a metric quantifying the cost reduction from utilizing 1~MWh of renewable electricity in a chemical's electrified process versus producing the same amount via its fossil-based process. We identify this metric as the driver behind the chemicals' order of transition, with methanol transitioning first and the aromatics last. Finally, we evaluate the interplay between the chemical industry and the overall energy system, uncovering the flexibility provision to the energy system that incentivizes an earlier transition of the chemical industry.

 In Section \ref{modeling}, we introduce the model setup used to represent the sector-coupled energy system together with the chemical industry. In Section \ref{results}, we present the results and discuss the findings. In Section \ref{conclusions}, we summarize our findings and highlight the key takeaways.

\section{Modeling the chemical industry transition pathway within a sector-coupled energy system}\label{modeling}

\noindent In this section, we introduce the modeling of the integrated sector-coupled energy system with the chemical industry to calculate the optimal transition pathway towards net-zero GHG emissions. Section~ \ref{chemicals} describes the representation of the chemical industry and its implementation within the sector-coupled energy system. Section~\ref{SCES} describes the representation of the German sector-coupled energy system used in our case study. Section~\ref{SecMOD} introduces SecMOD, the modeling framework used to calculate the transition pathways and describes the optimization setup details. Section~\ref{cost_avoided} introduces the \textit{Cost-Avoided} metric which drives the prioritization for electrification across the energy system.

\subsection{The chemical industry}\label{chemicals}

\begin{figure}
 \centering
 \includegraphics[width=0.6\textwidth]{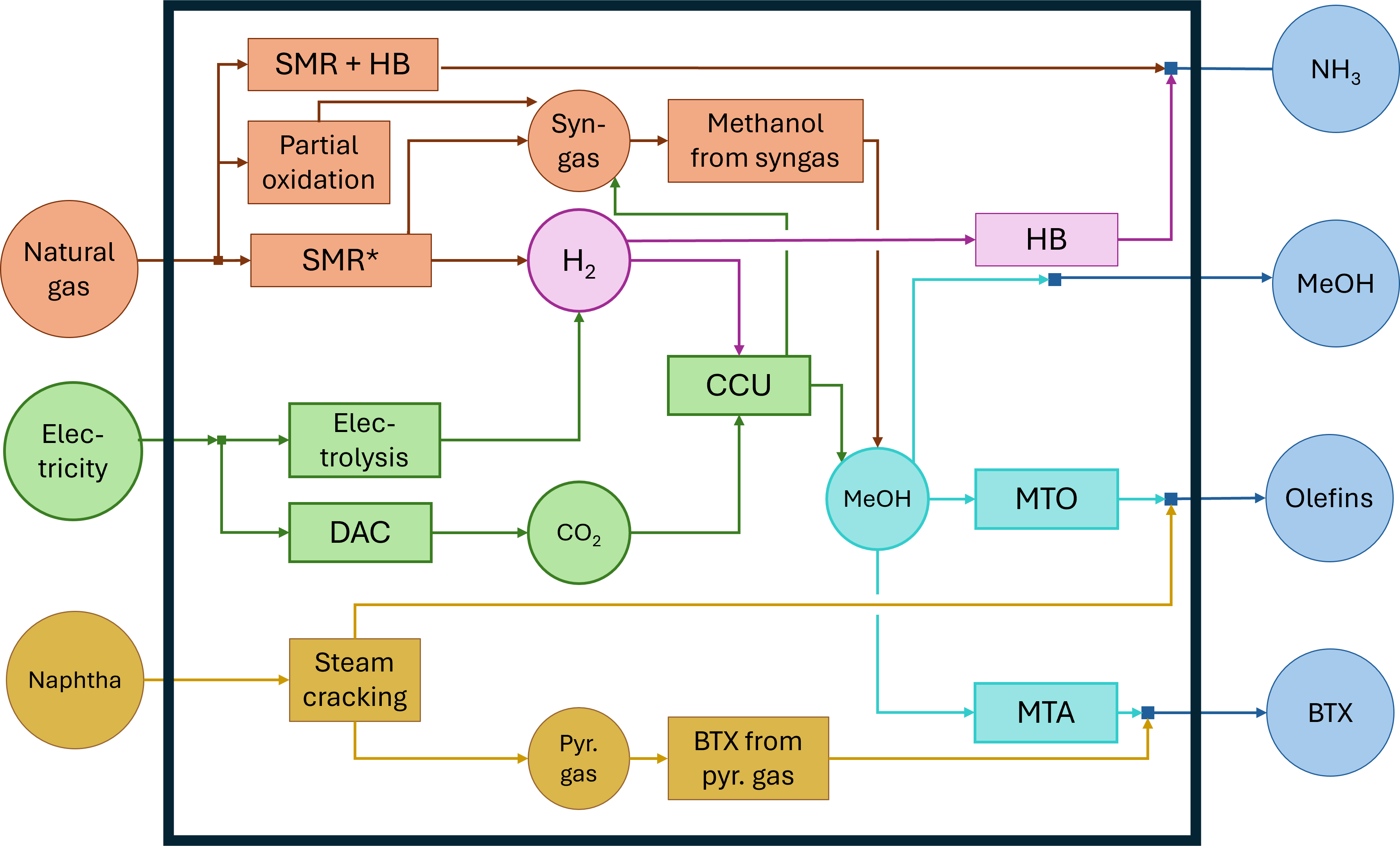}
 \caption{Schematic of processes included in the chemical industry model for producing base and high-value chemicals (blue boxes on the right). Olefins comprise ethylene and propylene, BTX comprise the aromatics benzene, toluene, xylene.  Processes are in boxes, whereas products are in circles. Processes are grouped by color based on the main product. SMR: steam methane reforming, HB: Haber Bosch, DAC: direct air capture, syngas: synthesis gas, pyr. gas: pyrolysis gas, NG: natural gas, MTO: methanol to olefins, MTA: methanol to aromatics, $\mathrm{NH_{3}}$: ammonia, $\mathrm{H_{2}}$: hydrogen, $\mathrm{CO_{2}}$: carbon dioxide, MeOH: methanol, CCU: carbon capture and utilization. *Three SMR processes are considered: one for $\mathrm{H_{2}}$ production, and two for syngas production. For syngas, SMR with $\mathrm{H_{2}}$ skimming and SMR with $\mathrm{CO_{2}}$ import are considered \cite{bachmann2023syngas}.}
 \label{fgr:schematic}
\end{figure}

Our representation of the chemical industry consists of the base and high value chemicals: ammonia, methanol, ethylene and propylene (olefins), and benzene, xylene, and toluene (aromatics). We introduce exogenous demands for the production of each chemical corresponding to historic German production volumes \cite{vci.}, and a constant hourly demand for every hour of the year. The yearly demands (\ref{tab:chemical processes}) are maintained constant throughout our transition pathway, as studies project constant or even declining chemical production volumes as countries transition towards carbon neutrality \cite{kloo2023towards, molecule.managers}. For each chemical, we include fossil-based and electrified production options (\ref{tab:chemical processes}), considering all process energy and material requirements. For olefins and aromatics electrified processes, we consider the methanol-to-olefins (MTO) and methanol-to-aromatics (MTA) processes. We do not consider the direct conversion of $\mathrm{CO_{2}}$ to olefins and aromatics due to their low technology readiness levels \cite{Lopez.2023}. Besides the main fossil-based and electrified processes for each chemical, we consider upstream processes for the production of chemical intermediates such as synthesis gas and pyrolysis gas (\ref{fgr:schematic}). In total, our chemical industry models over 30 processes gathered from a literature review (Supplementary Section~1).

\begin{table}
    \centering
    \caption{Yearly demand based on historic production volumes\cite{vci.}, fossil-based process and electrified process for each chemical included in our chemical industry model. SMR: steam methane reforming, HB: Haber-Bosch, $\mathrm{\textit{e-}H_{2}}$: electrolytic hydrogen, CCU: carbon capture and utilization, MTO: methanol-to-olefins, MTA: methanol-to-aromatics. *mixed xylenes}
    \begin{tabular}{llll}
        \toprule
        \textbf{Chemical} & \textbf{Fossil-based} & \textbf{Electrified} & \textbf{Demand}\\
          & \textbf{process} & \textbf{process}& $[\nicefrac{\mathrm{Mtonne}}{\mathrm{yr}}]$\\
         \midrule
         \vspace{0.2cm}
        ammonia  & SMR + HB & $\mathrm{\textit{e-}H_{2}}$ + HB& 2.56\\
        \vspace{0.2cm}
        methanol  & from synthesis gas & $\mathrm{\textit{e-}H_{2}}$ + CCU & 1.40\\
        ethylene  & steam cracking & \multirow{2}{*}{MTO} & 4.52\\
        \vspace{0.2cm}
        propylene & of naphtha  & & 3.44\\
        benzene  & solvent extraction& \multirow{3}{*}{MTA} & 1.51\\
        toluene  & from pyrolysis & & 0.55\\ 
        xylene* & gasoline & & 0.40\\
        \bottomrule
    \end{tabular}
    \label{tab:chemical processes}
\end{table}

The two key molecules needed for an electrified chemical industry are $\mathrm{CO_{2}}$ and $\mathrm{H_{2}}$. For the sourcing of $\mathrm{CO_{2}}$, we include direct air capture (DAC) \cite{Shu.2023} and industrial point source capture from the modelled chemical processes that separate a concentrated $\mathrm{CO_{2}}$ stream (i.e. $\mathrm{CO_{2}}$ from steam methane reforming). For the sourcing of $\mathrm{H_{2}}$, we include domestic production through steam methane reforming or electrolysis, and green $\mathrm{H_{2}}$ imports. To isolate the effect of inter-sectoral competition for limited renewable electricity, we place a high price penalty on imported green $\mathrm{H_{2}}$ such that the system prioritizes domestic energy resources.

Our chemical industry emissions are calculated as $\mathrm{CO_{2}}$-eq following the IPCC GWP-100 methodology \cite{edenhofer2015climate} for life cycle assessment. We consider the direct process emissions, life cycle emissions from the process energy requirements, life cycle emissions of imported products, and emissions from the chemicals use-phase and end-of-life. For the direct process emissions, we follow the methodology employed by Meys et al. \cite{meys2021achieving}, closing the atom balances around the chemical processes. Emissions from the process energy requirements are accounted for in the respective production technologies modelled in the energy system (Section \ref{SCES}). Emissions of imported products are taken from the ecoinvent database \cite{ecoinvent3.6}. For the chemicals use phase and end-of-life emissions, we follow the methodology employed by Zibunas et al. \cite{ZIBUNAS2022107798}, assuming complete combustion of the produced chemicals such that the carbon contained in the chemicals is converted to $\mathrm{CO_{2}}$ in the year that the chemicals are produced. We do not consider emissions associated with the construction of the chemical facilities. However, because we only constrain the system's operational emissions, in line with current carbon accounting practice \cite{change20062006}, neglecting the emissions associated with facility construction does not affect the results.

To account for existing fossil-based production capacities, we follow the methodology employed by Zibunas et al. \cite{ZIBUNAS2022107798}, assuming an existing capacity equal to each chemical's hourly demand, and a uniform age distribution for the facilities such that they retire uniformly throughout the transition pathway. This implementation results in an equal share of facilities retiring in each investment period, introducing the decision to either reinvest in fossil capacity, or replace it with electrified production capacity.

To achieve net-zero operational GHG emissions, the chemical industry in our model needs to fully electrify by 2045. Through this setup, we can address our research questions regarding the timing of the chemical industry's transition relative to the other energy sectors, as well as the interactions between a transitioning chemical industry and the energy system. 

\subsection{The Sector-Coupled Energy System}\label{SCES}
For our case study, we adopt the German energy system model developed by Baumgärtner et al. \cite{Baumgartner.2021} with the extensions implemented by Shu et al. \cite{Shu.2023}. The energy system considers the sectors: electricity, private mobility, residential heating, and industrial heating at three temperature levels (low temperature below 100°C, medium temperature heat between 100–400°C, and high temperature heat above 400°C). Sectoral demands are exogenously provided with an hourly resolution. Yearly emissions targets are also exogenously provided considering the historical emissions of the aforementioned sectors. The underlying assumptions can be found in the Supplementary Information of Baumgärtner et al. \cite{Baumgartner.2021}.

To include the chemical industry in the sector-coupled energy system model, we subtract the chemical industry electricity and heat demands from the original exogenous demands, as these demands are accounted for in the chemical processes (Supplementary Table~4 and Supplementary Table~5). We add the direct process emissions associated with the chemical industry \cite{FederalMinistryfortheEnvironmentNatureConservationandNuclearSafety.} and the use phase and end-of life emissions of the considered chemical products (Section~\ref{chemicals}) to the GHG emissions limit. We also update the exogenous emissions targets to reflect the most recent reduction targets of 65\% of 1990 levels by 2030 and net-zero by 2045 \cite{bundestag2021erstes} (Supplementary Table~6). We explicitly model $\mathrm{CO_{2}}$ as a product to account for $\mathrm{CO_{2}}$  production and consumption across various processes. Finally, we add resistance heaters and $\mathrm{H_{2}}$ boilers to introduce electrified high-temperature heat options. Due to the low technology readiness levels of high-temperature resistance heaters\cite{wismann2019electrified, qazi2022future, rieks2015experimental} and hydrogen boilers\cite{leicher2024electrification, lowes2023regret} for large-scale industrial applications, we assume high capital costs for these technologies (Supplementary Table~3). We include a scenario with optimistic capital cost assumptions in Supplementary Section~5.

Because we consider life cycle emissions calculated as $\mathrm{CO_{2}}$-eq \cite{edenhofer2015climate}, and we do not include $\mathrm{CO_{2}}$ sequestration in our system set-up, residual emissions are unavoidable. For instance, the life cycle emissions associated with the sorbent for direct air capture, cannot be offset when $\mathrm{CO_{2}}$ is captured and used as a feedstock. Aditionally, even electrified technologies that run on fully renewable electricity are subject to maintenance and degradation that would lead to life cycle emissions. To address this limitation and still reach our net-zero emissions target, a high price penalty is placed on remaining $\mathrm{CO_{2}}$-eq representing a high-cost $\mathrm{CO_{2}}$ removal option for achieving net-zero emissions.

\subsection{The SecMOD Framework and Optimization Setup}\label{SecMOD}
We employ the open-source, linear optimization framework SecMOD, used for sector-coupled energy system modeling, optimization, and life-cycle assessment (LCA) \cite{Reinert.2022}. The framework minimizes a user-defined objective function, such as cost minimization, subject to user-defined constraints, such as GHG emissions limits. SecMOD considers both spatial and temporal resolution, with the temporal resolution occurring at two levels: the time steps considered within the optimization of a single investment period, and the number of investment periods considered for a transition pathway optimization. For a single investment period, the full hourly time series is aggregated into user-defined typical periods using the TSAM package \cite{hoffmann2020review}. For the full transition pathway, the number of investment periods is defined by the user. Each investment period is optimized individually, with the user specifying the foresight regarding future periods. The investment periods can either be solved independently with no foresight of future periods, all together with perfect foresight of all investment periods, or with limited foresight by employing a rolling-horizon strategy \cite{Baumgartner.2021}. Further details of the modeling framework, including the mathematical formulation,  can be found in \cite{Reinert.2022} and in the open source repository \footnote{\url{https://git-ce.rwth-aachen.de/ltt/secmod}}.

In this study, SecMOD is used to calculate the cost-optimal transition pathway of the coupled German energy system (Section~\ref{SCES}) plus chemical industry (Section~\ref{chemicals}), subject to annual emissions constraints. We consider both investment and operating costs in our objective function, while only considering operating emissions in our constraint, in line with current accounting practice \cite{change20062006}. We solve the transition pathway for the years 2020 to 2045 in 5-year increments, instantiating the model for the base year 2016 \cite{Baumgartner.2021, Shu.2023}. We use a rolling horizon-strategy for the optimization of each investment period with a foresight of 4 periods, or 20 years. We represent the sector-coupled energy system with one node, and aggregate the hourly time series into 6 typical days of 6 hours each. The temporal resolution was taken from Shu et al. \cite{Shu.2023} By excluding spatial resolution in our system set-up, we potentially underestimate the energy system’s flexibility needs \cite{pfenninger2014energy} and disregard the spatial distribution of a transitioning chemical industry. However, as we are interested in the timing of the chemical industry's transition relative to other sectors, and we do consider temporal resolution, we believe that our setup is sufficient for our research objectives and leave the spatial component for future work. 

\subsection{\textit{Cost-Avoided} and the merit order curve: drivers of electrified production across the energy system}\label{cost_avoided}
The \textit{Cost-Avoided} by electrification, or $\Delta C_{i,t}^{elec}~\mathrm{[\frac{k\euro}{MWh}]}$, quantifies the system's cost reduction per MWh of renewable electricity used in the electrified production of product $i$ compared to producing an equivalent amount via its fossil-based alternative (Equation~\ref{delC_i}). This \textit{Cost-Avoided} can be interpreted as an economic Power-to-X efficiency following the methodology introduced by \citet{sternberg2015power}. The \textit{Cost-Avoided} depends on time, $t$, based on the time-dependent operation of the sector-coupled energy system. A positive \textit{Cost-Avoided} indicates a decrease in the system costs.

    \begin{equation}
    \Delta C_{i,t}^{elec}~=~\Delta C_{i,t}^{op}~+~(C^{CO_{2}}_{y} \cdot~\Delta e_{i,t}^{CO_{2}})
    \label{delC_i}
    \end{equation}

The \textit{Cost-Avoided} is comprised of three parts: 
\begin{enumerate}
    \item $\Delta C_{i,t}^{op}~\mathrm{[\frac{k\euro}{MWh}]}$: the difference in operating costs between 1~MWh-worth of product $i$ produced via its electrified process versus producing the same amount via its fossil-based process at time $t$ (Equation~\ref{delC_op}):

    \begin{equation}
    \Delta C_{i,t}^{op}~=~M_{i,t}~\cdot~(C_{i,t}^{op,fossil}~-~C_{i,t}^{op,elec})
    \label{delC_op}
    \end{equation}

    where $M_{i,t}~\mathrm{[\frac{unit_{i}}{MWh}]}$ is the amount of electrified product $i$ produced with 1~MWh of renewable electricity at time $t$, $C_{i,t}^{op,elec}~\mathrm{[\frac{k\euro}{unit_{i}}]}$ and $C_{i,t}^{op,fossil}~\mathrm{[\frac{k\euro}{unit_{i}}]}$ are the operating costs of the electrified and fossil-based processes per amount of product $i$ at time $t$, respectively. The amount of product $i$, $[\mathrm{unit_{i}}]$ can correspond to tonne, MWh, or vehicle~km depending on the product, $i$.
    
    \item $\Delta e_{i,t}^{CO_{2}}~\mathrm{[\frac{tonne~CO_{2}\text{-}eq}{MWh}]}$: the difference in $\mathrm{CO_{2}}$-eq emissions between 1~MWh-worth of product $i$ produced via its electrified process versus producing the same amount via a fossil-based process at time $t$ (Equation~\ref{dele_op}). This calculation is based on \citet{sternberg2015power} with an added time component.

    \begin{equation}
    \Delta e_{i,t}^{CO_{2}}~=~M_{i,t}~\cdot~(e_{i,t}^{CO_{2},fossil}~-~e_{i,t}^{CO_{2},elec})
    \label{dele_op}
    \end{equation}

    where $M_{i,t}~\mathrm{[\frac{unit_{i}}{MWh}]}$ is the same as above. $e_{i,t}^{CO_{2},elec}~\mathrm{[\frac{tonne~CO_{2}\text{-}eq}{unit_{i}}]}$ and $e_{i,t}^{CO_{2},fossil}~\mathrm{[\frac{tonne~CO_{2}\text{-}eq}{unit_{i}}]}$ are the emissions of the electrified and fossil-based processes per amount of product $i$ at time $t$, respectively.
    
    \item $C^{CO_{2}}_y~\mathrm{[\frac{k\euro}{tonne~CO_{2}-eq}]}$: the energy system's $\mathrm{CO_{2}}$ price, represented by the endogenous shadow price of the optimization model's $\mathrm{CO_{2}}$-eq emissions constraint.  We obtain one $\mathrm{CO_{2}}$ price per investment period, $y$, from our total annual emissions constraint.
\end{enumerate}

To calculate $M_{i,t}$, we take the inverse of the electricity demand per unit of a product's electrified production at time $t$, $E_{i,t}~\mathrm{[\frac{MWh}{unit_{i}}]}$ (Equation~\ref{M_i}): 

 \begin{equation}
    M_{i,t} = (E_{i,t})^{-1}
    \label{M_i}
\end{equation}
 
The time component is introduced into the electricity demands ($E_{i,t}$), operating costs ($C_{i,t}^{op,elec}$, $C_{i,t}^{op,fossil}$) and emissions ($e_{i,t}^{CO_{2},elec}$, $e_{i,t}^{CO_{2},fossil}$) because these terms consider the direct production processes and the underlying supply chains of the process material and energy inputs (Supplementary Figure~1). The underlying supply chains depend on the temporal results of our optimization model. For example, let's consider electrified methanol production (using $\mathrm{H_{2}}$ from electrolysis and captured $\mathrm{CO_{2}}$) which requires medium-temperature heat. This heat can be produced via fossil fuel combustion depending on the endogenously optimized heat supply mix of the energy system at a given time-step. The costs and emissions for the underlying heat production are thus included in the costs, $C_{i,t}^{op,elec}$, and emissions, $e_{i,t}^{CO_{2},elec}$, of the electrified methanol process for the given time-step (Supplementary Section~3).

\begin{figure}[H]
 \centering
 \includegraphics[width=0.6\textwidth]{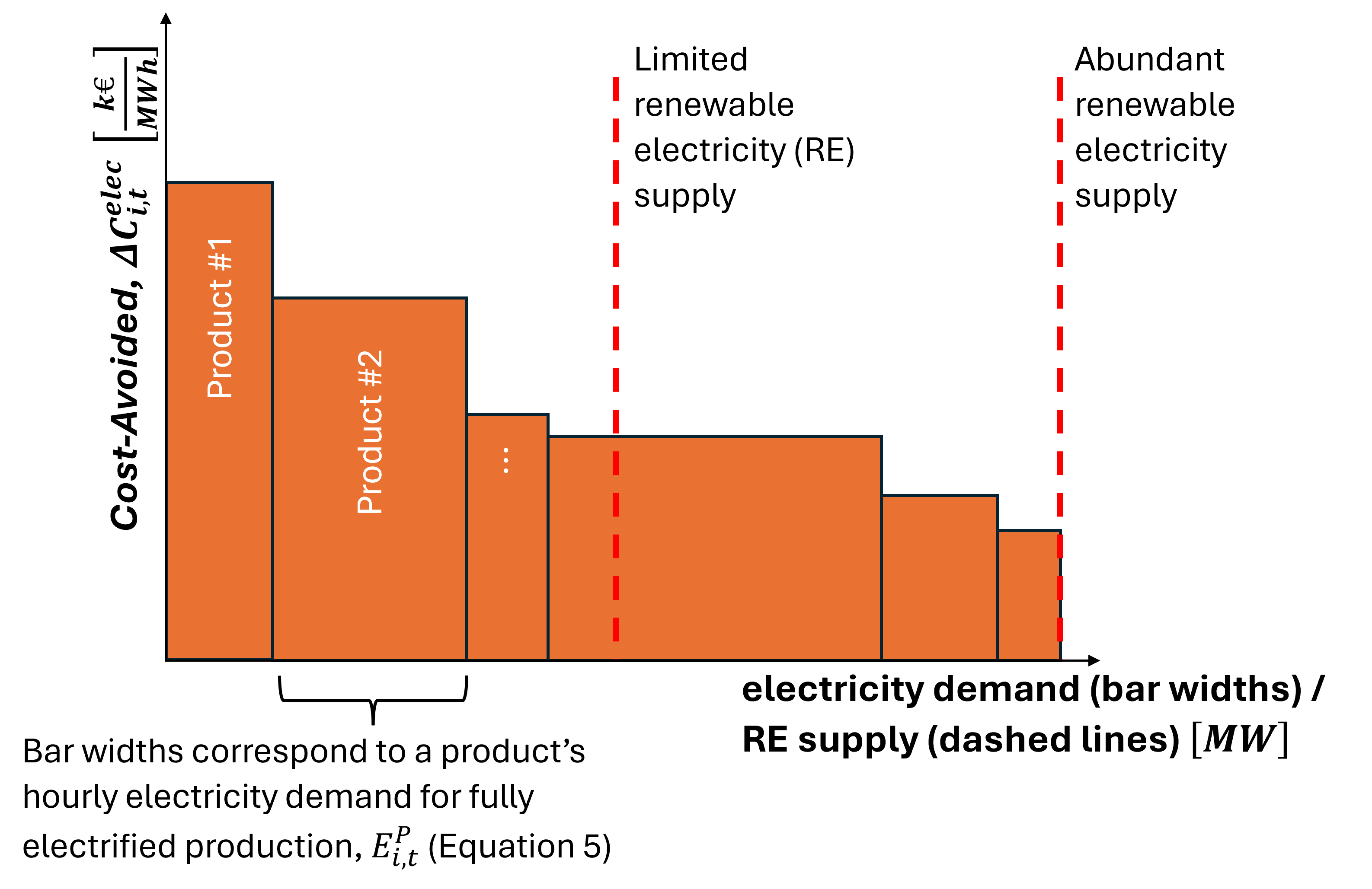}
 \caption{Schematic representation of the time-dependent merit order curve arising from the \textit{Cost-Avoided} of each product, $\Delta C_{i,t}^{elec}$, and their electricity demands, $E_{i,t}^{P}$. Here, for explanatory purposes, we show the intersection of two possible renewable electricity (RE) availabilities for a single merit order curve. However, a distinct merit order curve exists in each time step.} 
 \label{fgr:merit_order}
\end{figure}

 The \textit{Cost-Avoided} creates a merit order of electrified products, ranking products by their cost savings from electrified production. We combine this merit order with the products' hourly electricity demands, $E_{i,t}^{P}~\mathrm{[\frac{MWh}{hour}]}$ (Equation~\ref{E_IC_i}), to create a time-dependent merit order curve. Based on its intersection with the energy system's availability of renewable electricity at a given time-step, this curve determines which products are produced electrically and to what extent (\ref{fgr:merit_order}). This merit order curve complements the electricity-market merit order and is induced through the emissions constraint. The \textit{Cost-Avoided} and the resulting merit order curve therefore dictate the hourly production of electrified products in the energy system. 
 
 The products' hourly electricity demands for fully electrified production, $E_{i,t}^{P}~\mathrm{[\frac{MWh}{hour}]}$, are used as the bar widths for the merit order curve (Equation~\ref{E_IC_i}). These electricity demands are calculated from the minimum of a product's electrified installed capacity in investment year, $y$, $P_{i,y}~\mathrm{[\frac{tonne}{hour}]}$ and a product's exogenously provided hourly demand, $D_{i,t}~\mathrm{[\frac{tonne}{hour}]}$. We take the minimum since some products, such as methanol, serve as intermediates. Taking the minimum ensures that the electricity demand of the additional installed capacity for intermediary production is allocated to the final product rather than to the intermediate.

\begin{equation}
    E_{i,t}^{P} = E_{i,t}\cdot min(P_{i,y}, D_{i,t})
    \label{E_IC_i}
\end{equation}
 
 The \textit{Cost-Avoided} and the corresponding merit order curve provide a tool for evaluating the hourly deployment of electrified products across the energy system given renewables intermittency. This tool is particularly useful for determining the prioritization of products for electrification and for calculating and comparing utilization rates across the various electrified products. These rates can be calculated by aggregating the hourly deployment provided by the merit order curve. Furthermore, products positioned further down in the merit order can become valuable flexibility providers to the energy system by adapting their production between electrified and fossil-based depending on the availability of renewable electricity (\ref{fgr:merit_order}). Thus, the merit order curve is also a valuable tool for evaluating the flexibility provision from individual products or energy sectors. 

\section{Results and Discussion}\label{results}
\noindent Here, we present the results and discuss the findings of the coupled energy system and chemical industry's transition to net-zero emissions. We first focus on the overall system's transition pathway in Section~\ref{pathway}. We then take a deep dive into the chemical industry's transition in Section~\ref{chemicals transition}. In Section~\ref{flexibility}, we focus on the interactions between a transitioning chemical industry and energy system with particular emphasis on flexibility provision. 

\subsection{Transition pathway of the coupled energy system and chemical industry}\label{pathway}
\begin{figure}[ht!]
 \centering
 \includegraphics[width=0.6\textwidth]{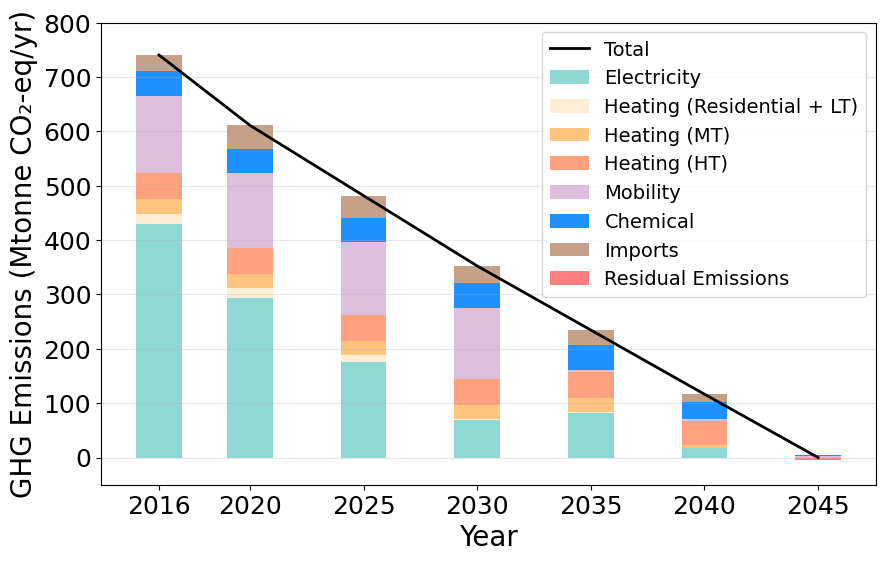}
 \caption{GHG emissions, in million tonnes $\mathrm{CO_{2}}$-eq, of the integrated energy system and chemical industry model by sector. The chemical industry transitions last along with medium temperature (MT) and high temperature (HT) heat starting in 2040. LT: low temperature.} 
 \label{fgr:transition_pathway}
\end{figure}

In its combined transition pathway with the energy system, the chemical industry starts implementing electrified production in 2040, after the transitions of the electricity, residential and low-temperature heat, and mobility sectors (\ref{fgr:transition_pathway}). The chemical industry starts transitioning together with medium and high-temperature heat. Hence, in energy systems with limited renewable electricity, priority is first placed on building up renewable electricity, transitioning to heat pumps and to battery electric vehicles before electrifying chemical production. Electrification of chemical production takes place along with the electrification of other hard-to-abate, electricity-intensive sectors. 

To enable the transition to net-zero emissions, the electricity sector first transitions away from lignite and coal, building up wind and photovoltaic capacities (Supplementary Figure~2). Electricity from natural gas combined cycle is used for dispatchable electricity between 2025 and 2040, causing most emissions from the electricity sector in that time period. The residential and low temperature heat sectors also transition from natural gas boilers to heat pumps together with the electricity sector (Supplementary Figure~3 and Supplementary Figure~4), taking advantage of the increasing renewable electricity availability. Next, the mobility sector transitions from a mix of gasoline, diesel, and natural gas cars to battery electric vehicles in 2035 (Supplementary Figure~5). The transition of the mobility sector occurs in one investment period due to the existing vehicle fleet reaching the end of its lifetime, and the foresight to the net-zero emissions target in 2045 influencing the decision to invest in electric vehicles rather than reinvest in fossil fuel vehicles. Finally, the chemical industry starts transitioning to electrified production in 2040 together with medium and high temperature heat. In 2040, 77\% of medium temperature heat is produced via electrode boilers rather than natural gas boilers (Supplementary Figure~6), whereas only 0.5\% of high temperature heat is produced via resistance heaters (Supplementary Figure~7). The bulk of the high temperature heat transition occurs in 2045, when 90\% of high temperature heat is produced via resistance heaters and $\mathrm{H_{2}}$ boilers and the remainder is produced as a by-product of an electrified chemical industry.

The late transition of the high temperature heat sector is driven in part by the high capital costs assumed for resistance heaters and $\mathrm{H_{2}}$ boilers (Section~\ref{SCES}). In a scenario with more optimistic cost assumptions (Supplementary Section~5), the high-temperature heat sector still transitions late, but the bulk occurs in 2040 rather than 2045. In this scenario, a portion of the chemical industry transition is delayed to 2045. This finding highlights the delayed transition of the chemical industry compared to other energy sectors, and underscores how closely its timeline is tied to the transition of other hard-to-abate, electricity-intensive sectors.

Despite the net-zero emissions target in 2045, 4.2~Mtonne of residual $\mathrm{CO_{2}}$-eq emissions remain. Although small, equivalent to only 0.7\% of Germany's emissions in 2020 \cite{InternationalEnergyAgency.Germany.2020}, the residual emissions indicate that a fully net-zero energy system is not possible without carbon dioxide removal. 75\% of these residual emissions come from the operation of battery electric vehicles (BEVs). These operational emissions come from our allocation of road and vehicle degradation to the operational life-cycle of BEVs, which we adopt from Baumgärtner et al.\cite{Baumgartner.2021}. 20\% of the residual emissions come from the chemical industry, mainly from the supply chain of the direct air capture units. The remaining residual emissions come from the operation of renewable electricity production technologies and from the $\mathrm{H_{2}}$ import supply chain.

To reach the nearly net-zero energy system in 2045 with a fully electrified chemical industry, the energy system imports 483~TWh/yr of clean energy in the form of green $\mathrm{H_{2}}$, equivalent to 41\% of the energy system's electricity demand. These imports are needed due to insufficient domestic renewable electricity availability during some hours of the year. 43\% of these imports are required for a fully electrified chemical industry either as direct $\mathrm{H_{2}}$ feedstock or for process energy. Although these imports are lower than present-day fossil-based energy imports to Germany, which imported 968~TWh of natural gas alone in 2023 \cite{Bundesnetzagentur.}, they require a seven-fold increase in present-day global green hydrogen production \cite{tonelli2023global} by 2045, which is a matter with much uncertainty \cite{odenweller2022probabilistic}. Hence, fully electrifying the chemical industry in countries with limited renewable electricity availability requires green energy imports of magnitudes which may not be available.

An interesting intermediate configuration, however, can be seen in 2040, when the chemical industry is partially electrified together with fossil-based production. The energy system imports 400~TWh of fossil-based energy, 33\% of which is used in the chemical industry as naphtha feedstock. No clean energy is imported due to the high cost assumption. The reduced reliance on clean energy imports from a mixed chemical industry suggests that combining electrified production with other production options can yield a more resilient energy system to global green energy availability. 

In summary, our study shows that the chemical industry transitions towards the end of the pathway, following the build-up of renewable electricity, the transition to heat pumps, and to battery electric vehicles. We find that a fully net-zero energy system is not possible, requiring carbon dioxide removal to offset residual emissions. Finally, a nearly net-zero system with a fully electrified chemical industry requires substantial clean energy imports, largely due to the chemical industry's high energy requirements. However, although the required energy imports are much lower than present-day fossil imports, our intermediate 2040 results show that electrified chemical production can potentially be combined with other production options to reduce import dependencies.

\subsection{Transition towards an electrified chemical industry}\label{chemicals transition}
\begin{figure}[H]
 \centering
 \includegraphics[width=0.5\linewidth]{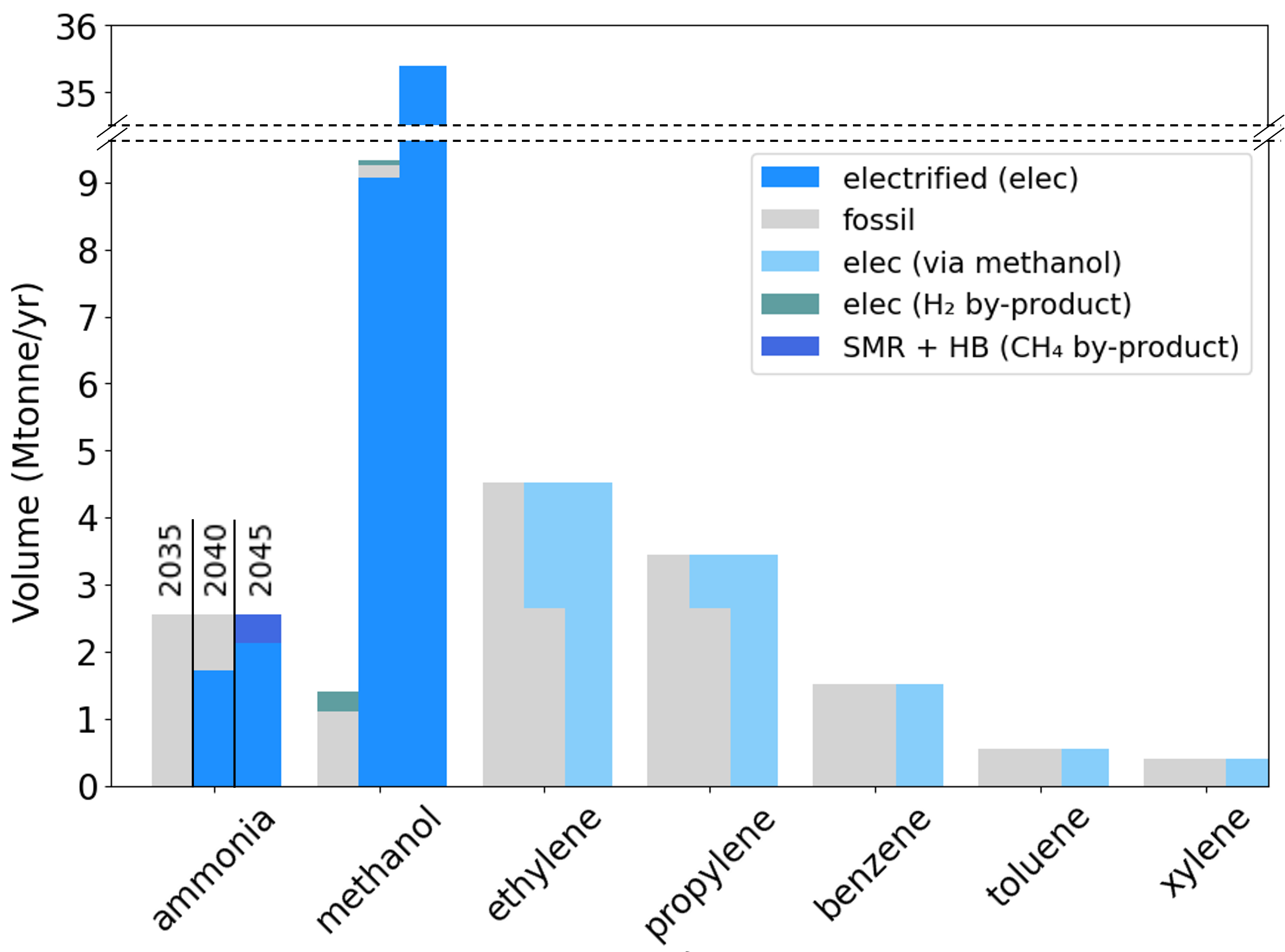}
 \caption{Transition of chemical production from 2035 to 2045. Years prior to 2035 have the same production mix as 2035 and are therefore excluded from the figure. Methanol becomes an important intermediate for the production of electrified olefins and aromatics, leading to a 25-fold increase in methanol production between 2035 and 2045. The specific electrified, fossil, and methanol-based processes are found in \ref{tab:chemical processes}. elec (via methanol) refers to the methanol-to-olefins and methanol-to-aromatics processes. elec ($\mathrm{H_{2}}$ by-product) refers to methanol produced via  CCU using by-product $\mathrm{H_{2}}$ from synthesis gas production and $\mathrm{CO_{2}}$ from chemical industry point sources. SMR + HB ($\mathrm{CH_{4}}$ by-product) refers to ammonia produced via  steam methane reforming + Haber-Bosch using by-product $\mathrm{CH_{4}}$ from other electrified processes.  } 
 \label{fgr:chem_ind_transition}
\end{figure}

Electrified chemical production begins in the year 2040, when methanol, ammonia, and the olefins (ethylene and propylene) are partly produced via their electrified processes. Their yearly production mix is split between fossil-based and electrified production (\ref{fgr:chem_ind_transition}). Before this transition year, all chemicals are produced via their fossil-based processes except for 20\% of methanol, which is  produced via CCU by combining by-product $\mathrm{H_{2}}$ from synthesis gas production with $\mathrm{CO_{2}}$ from chemical industry point sources. The aromatics (benzene, toluene, and xylene) transition at the end of the pathway in 2045 when all chemicals are fully produced electrically to meet the net-zero emissions target.

The need for methanol as an intermediate for electrified production of olefins and aromatics increases methanol production drastically by 2045. In a fully electrified chemical industry, methanol production increases 25-fold compared to levels before 2040. This increase indicates the need for a massive scale-up of methanol production to transition to a fully electrified chemical industry when relying on high TRL methanol-to-olefins and methanol-to-aromatics processes. 

The chemicals' order of transition can be understood by the merit order of chemicals created by the \textit{Cost-Avoided} (Section~\ref{cost_avoided}). In the transition year, 2040, the merit order stays the same for every time-step despite the time dependency of the \textit{Cost-Avoided}, with methanol consistently first, followed by ammonia, the olefins, and finally the aromatics  (\ref{tab:cost_avoided_breakdown}). The resulting order is attributed to two main components: the electricity requirements for the chemicals' electrified production ($E_{i,t}$), and the emissions-intensity of their respective fossil-based processes ($e_{i,t}^{CO_{2},fossil}$). Methanol has the highest \textit{Cost-Avoided}, despite the higher electricity requirement per tonne of production than for ammonia (\ref{tab:cost_avoided_breakdown}). Methanol's lead is due to the high emissions associated with its fossil-based production, resulting in a high $\Delta e_{methanol}^{CO_{2}}$ (Equation~\ref{dele_op}). The emissions abated via methanol's electrified production are high enough that a smaller mass of electrified methanol production abates more emissions than a larger mass of ammonia replacing its fossil-based process. The placement of the aromatics as last in the merit order is due to their high electricity requirement, particularly for production of methanol as a feedstock.

Overall, our results indicate an order of transition for the individual chemicals, with methanol first, followed by ammonia, the olefins, and finally the aromatics. A fully electrified chemical industry would require a massive scale-up of methanol production, requiring 25 times more production than today. The order of transition is driven by the merit order of the \textit{Cost-Avoided}, with a chemical’s position in the merit order dictated by both its electricity demand for electrified production, and the emissions-intensity of its fossil-based alternative processes. Hence, prioritization of chemicals for electrified production should consider both their electricity requirements, and the emissions intensity of their fossil-based alternatives.

While the merit order drives the order of transition, an interplay between the chemical industry and the energy system orchestrates the partial electrification in the transition year. We explore this interplay in the following section (Section~\ref{flexibility}).

\begin{table}[H]
    \centering
    \small
    \renewcommand{\arraystretch}{1.5} 
    \caption{Average \textit{Cost-Avoided} ($\Delta C_{i,elec}$) (Equation~\ref{delC_i}), and components of \textit{Cost-Avoided}: electricity requirements ($E_{i}$), $\Delta e_{i}^{CO_{2}}$, $\Delta C_{i}^{op}$, and $C^{CO_{2}}$, for the year 2040 using the aggregated production mix as the underlying supply chain. Olefins represent both ethylene and propylene, which have the same values. Aromatics represent benzene, toluene, and xylene, which have the same values. Due to the low $\Delta C_{i}$ of the aromatics, no installed electrified capacities exist in 2040.}
    \label{tab:cost_avoided_breakdown}
    \begin{tabularx}{0.8\linewidth}{l|XXXXX}
        \toprule
          & \textbf{Unit} & \textbf{Methanol} & \textbf{Ammonia} & \textbf{Olefins} & \textbf{Aromatics} \\
        \midrule
        $\mathbf{\Delta C_{i}}$ & $\mathrm{\frac{k\euro}{MWh}}$ & 1.05 & 0.44 & 0.41 & 0.19 \\
        $\mathbf{E_{i}}$ & $\mathrm{\frac{MWh}{tonne~i}}$ & 11.1 & 8.8 & 34.3 & 52.0 \\
        $\mathbf{\Delta e_{i}^{CO_{2}}}$ & $\mathrm{\frac{tonne~CO_{2}-eq}{tonne~i}}$ & 0.33 & 0.11 & 0.14 & 0.07 \\
        $\mathbf{\Delta C_{i}^{op}}$ & $\mathrm{\frac{k\euro}{tonne~i}}$ & 0.24 & 0.17 & 0.07 & 0.02 \\
        $\mathbf{C^{CO_{2}}}$ & $\mathrm{\frac{k\euro}{tonne~CO_{2}-eq}}$ & \multicolumn{4}{c}{2.46} \\
        \bottomrule
    \end{tabularx}
\end{table}


\subsection{Flexibility provision from a transitioning chemical industry}\label{flexibility}

The partial electrification of multiple chemicals observed in 2040 (Section~\ref{chemicals transition}) contrasts previous studies that indicate full sequential transitions of individual chemicals \cite{Katelhon.2019, ZIBUNAS2022107798, Rixhon.2022}. However, these studies do not consider the interactions with the energy system. By considering these interactions, we find that the partial electrification is beneficial due to flexibility that the chemical industry provides to the energy system once investments in both fossil-based and electrified production capacities are made.

In the transition year, 2040, which still has a positive emissions budget, the system re-invests in the phased-out fossil-based production capacities to meet the full hourly demand (Section~\ref{chemicals}), and, at the same time, invests in electrified production capacities to meet between 50\% (ethylene) to 100\% (methanol, ammonia, propylene) of the hourly demand. These investments lead to an over-sizing of the chemicals' installed capacities and to diversification in production options. Therefore, in a given hour, chemicals can either be produced from fossil-based or electrified processes depending on the renewable electricity supply and on the renewable electricity demand from the other sectors. The ability to choose between chemical production options thus introduces a flexibility lever to the energy system.

\begin{figure}[H]
 \centering
 \includegraphics[width=0.9\textwidth]{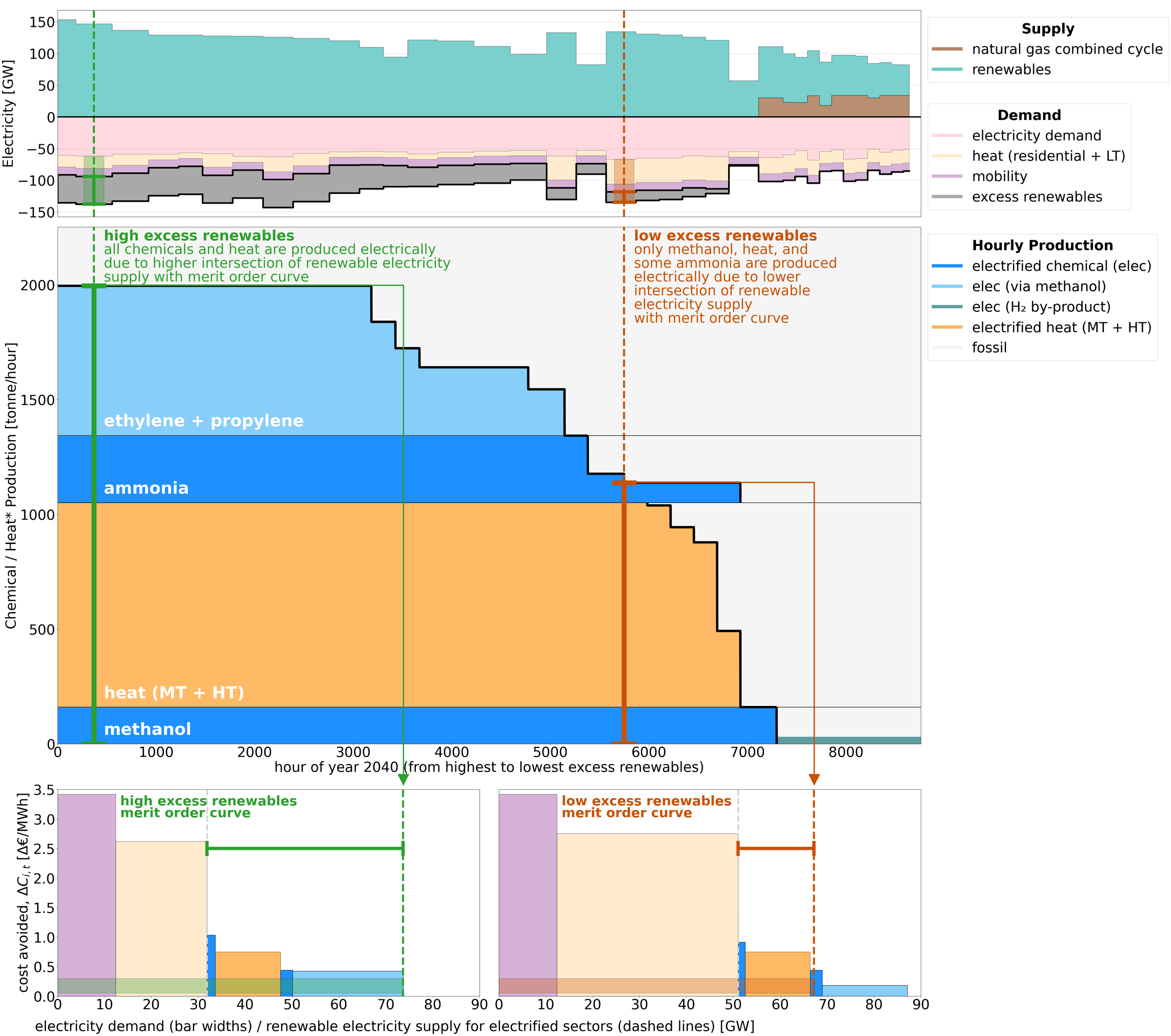}
 \caption{\textbf{TOP}: electricity supply and demand for each hour in the year 2040. The hours are ordered from highest to lowest excess renewables after full electrification of the electricity, residential and low-temperature (LT) heat, and mobility sectors. Due to the time series aggregation, the hours repeat themselves, causing the steps in the figure.  Electricity storage is excluded from the figure. \textbf{MIDDLE}: load-duration curve for the year 2040, with the hourly breakdown of electrified vs. fossil-based production for medium and high-temperature (MT + HT) heat and for each chemical product. The hours are in the same order as in the electricity balance plot (top figure). In hours with high excess renewables (dashed green line), all chemicals and heat are produced via their electrified process up to the installed capacities. In hours with low excess renewables (dashed red line), only a subset of chemicals are produced via their electrified processes. This behavior is explained by the merit order curves created by the \textit{Cost-Avoided} and the electricity demand of each product (bottom figures). \textbf{BOTTOM}:  merit order curves of electrified products in the sector-coupled energy system. Each curve corresponds to a separate hour, identified by the red and green dashed lines crossing the top and middle figures. The red and green dashed lines show the renewable electricity supplied for the electrified energy sectors. Everything to the left of the intersection between the renewable electricity supply and the merit order curve is produced electrically for that hour. Mobility, residential heat, and LT heat are fully electrified in every hour (dashed gray lines). The excess renewables (red and green brackets) are then used for electrification of chemicals and MT + HT heat. There is a tranche of electrified ammonia that is prioritized over electrified MT~+~HT heat in the middle figure while having a lower \textit{Cost-Avoided}. Although not shown in this figure, the \textit{Cost-Avoided} of ammonia can be split into two parts depending on whether the point-source $\mathrm{CO_{2}}$ emissions from fossil-based ammonia can be used downstream to produce CCU-based methanol (Supplementary Section~3) Thus, the prioritized tranche of electrified ammonia corresponds to the portion with a higher \textit{Cost-Avoided}. *Heat production is shown in tonne natural gas equivalents using a heating value of $\mathrm{15.4~\frac{MWh}{tonne}}$.} 
 \label{fgr:flexibility_provision}
\end{figure}

The mechanism for flexibility provision is depicted in \ref{fgr:flexibility_provision}: renewable electricity supply is prioritized in the electricity, residential and low-temperature heat, and mobility sectors due to their higher \textit{Cost-Avoided} (Section~\ref{cost_avoided}). These sectors are always 100\% electrified. The remaining renewable electricity we thus consider as excess renewables that can then be used by the medium to high-temperature heat and the chemical sectors (\ref{fgr:flexibility_provision}, TOP). The merit order curve, (Section~\ref{cost_avoided}), dictates the electrified portion of these additional sectors in a given hour based on the curve's intersection with the renewable electricity supply (\ref{fgr:flexibility_provision}, BOTTOM). In hours with abundant excess renewables, all of the medium to high-temperature heat and chemicals are produced electrically up to the electrified installed capacities (\ref{fgr:flexibility_provision}, dashed green line). However, in hours with limited excess renewables, only a portion of these sectors is fulfilled with electrified production, with the remainder produced via fossil-based processes.  This switching between fossil-based and electrified production in the individual hours of the year leads to the partial electrification of chemicals over the course of the year 2040. 
 
A closer look at the hourly chemical production mix for 2040  (\ref{fgr:flexibility_provision}, MIDDLE) shows that sufficient renewable electricity is available 80\% of the time to fully produce electrified methanol, which is highest in the merit order following low-temperature heat, while sufficient renewables are only available 34\% of the time for maximum electrified production of olefins at the back of the merit order. The large variation in the \textit{Cost-Avoided} across the different chemicals together with the chemical industry's large electricity demand make the chemical sector span a large portion of the merit order curve, introducing a large flexibility lever to the energy system. This flexibility lever makes investments in electrified chemical production capacities cost-optimal despite full-load utilization rates as low as 34\%.

Our results reveal that over-sizing and diversifying the chemical industry with electrified production capacities provides valuable flexibility to a renewables-dominated energy system. This flexibility provision makes earlier investments in electrified capacities advantageous despite lower utilization rates. The system-wide benefits of over-sizing the chemical industry could serve as an opportunity for the industry to accelerate its transition to electrified production.

Implementing this flexibility would require companies to invest in multiple chemical production capacities in parallel, which may seem challenging given the decreased utilization and prolonged payback periods that do not capture the economic benefits to the energy system. However, companies can benefit from these investments through market dynamics, such as fluctuating electricity and feedstock prices, and through risk mitigation, such as disruptions in supply chains. In fact, studies already show that making electrified chemical production flexible to market dynamics, either via process diversification \cite{laky2024market} or through over-capacities \cite{guerra2023barriers}, can yield economic gains despite lower utilization rates. Guerra et al. \cite{guerra2023barriers} demonstrate this phenomenon for electrolyzer operation, showing how lower utilization rates translate to lower electricity prices for the electrolyzer.

Furthermore, for countries to obtain the system-wide benefits of an over-sized and diversified chemical industry, governments can put financial incentives in place for construction of parallel production capacities, sharing some of the economic benefits with companies. Incentives for the deployment of green industry can already be seen, for example, through the Inflation Reduction Act \cite{bistline2023emissions} and the European Green Deal \cite{fetting2020european}, manifesting the feasibility of such incentives. Hence, benefits at both the country and the company level may be sufficient to incentivize the roll-out of an over-sized and diversified chemical industry.

\section{Conclusions}\label{conclusions}
In this work, we investigate the chemical industry's pathway to electrified production within the context of a sector-coupled national energy system's transition to net-zero emissions. We determine the timing of the chemical industry's transition to electrified production relative to other sectors, and resolve the interactions between a transitioning chemical industry and the energy system.

Our results show that the build-up of renewable electricity, the transition to heat pumps for low-temperature heat and to battery electric vehicles for mobility are prioritized over electrified chemical production. We find that a fully net-zero energy system requires $\mathrm{473~\frac{TWh}{yr}}$ of green energy imports, equivalent to 41\% of its annual electricity demand, with 43\% of these imports used by a fully electrified chemical industry. Although much lower than present day fossil-based energy imports, these imports require a massive scale-up in global green energy production. Furthermore, full chemical electrification requires a 25-fold increase in annual methanol production compared to today when considering high TRL methanol-to-olefins and methanol-to-aromatics electrified processes.
 
Green energy imports and methanol production can be reduced via partial electrification of the chemical industry. We find that methanol, ammonia, and the olefins start transitioning to electrified production together with the medium and high-temperature heat sectors with lower import dependencies. These chemicals are distinguished by their high \textit{Cost-Avoided}, a metric quantifying the system cost reduction per MWh of renewable electricity used in their electrified processes. Of our considered chemicals, methanol has the highest \textit{Cost-Avoided} due to the high emissions of its fossil-based process, which its electrified process abates. Aromatics have the lowest \textit{Cost-Avoided} due to their high electricity requirement, particularly for production of methanol as a feedstock. Thus, prime targets for electrification are chemicals with lower electricity requirements or high-emission fossil-based alternatives, such as methanol and ammonia. 

 Our results expose an additional benefit of partial electrification of the chemical industry: flexibility provision to the energy system. Methanol, ammonia, and the olefins provide flexibility when both fossil-based and electrified production capacities are available. Production of these chemicals adapts to the hourly availability of renewable electricity by adjusting the production mix accordingly. This flexibility provision makes earlier investments in electrified capacities worthwhile despite full-load utilization rates as low as 34\%. The energy system benefits of a partially electrified chemical industry can potentially be shared with individual companies in the form of subsidies or tax credits to incentivize an earlier roll-out of electrified chemical production capacities.

To more holistically determine the role of electrified chemical production, it should be evaluated together with other sustainable chemical industry alternatives like biomass, recycling, and CCS. A dedicated consideration of green hydrogen imports would also be beneficial, as these imports can influence an electrified chemical industry's demand for domestic renewable electricity as well as the energy system's flexibility requirements. Nonetheless, our study shows that electrified chemical production could be a valuable part of the transition to net-zero emissions if prioritized after other energy sectors. Building up electrified production capacities while diversifying with other dispatchable production options provides both an avenue to defossilization of chemical production and flexibility to the energy system, thus serving as a valuable component in net-zero energy systems.

\section*{Acknowledgements}
This work was funded by the Swiss Federal Office of Energy’s SWEET program as part of the project PATHFNDR.

\clearpage

\bibliography{bibliography} 
\bibliographystyle{bibliography} 

\end{document}


\maketitle

\tableofcontents

\section{Chemical industry process data}
\label{sec:SI-data}
In this section, we provide data for the processes included in our chemical industry model. We provide specific data where publicly available. Process flow data is provided in \ref{tab:SI-flows}. The functional unit for each process is indicated by the flow with the value of 1. Negative flows indicate process inputs. IHS refers to the IHS Process Economics Program \cite{IHS} which provides process flows and economic data for chemical facilities. \ref{tab:SI-temp_levels} provides the temperature (T) ranges for low, medium, and high T heat from Baumgärtner et al.\cite{Baumgartner.2021}.

\ref{tab:SI-costs_emissions} provides data on process costs and net GHG emissions calculated as $\mathrm{CO_{2}}$-eq following the IPCC GWP-100 methodology \cite{edenhofer2015climate}. The costs and emissions are reported per process functional unit reported in \ref{tab:SI-flows}. Flows, costs, and emissions differentiated by year indicate a projected technological evolution throughout the transition pathway. Investment costs are annualized using the methodology in Baumgärtner et al.\cite{Baumgartner.2021}, using an interest rate of 5\% and payback period of 30~years for the chemical facilities. 

\newpage
\renewcommand{\arraystretch}{1.3}
\begin{longtable}{ll|llll}
\captionsetup{justification=centering, width=\textwidth}
\caption{Process data for chemical industry model}
    \label{tab:SI-flows}\\
    \centering
         \textbf{Product} & \textbf{Process} & \multicolumn{3}{l}{\textbf{Flows}} & \\
         \specialrule{.2em}{0em}{0em} 
         ammonia & Haber-Bosch & ammonia & 1 & tonne/hr & \cite{mayer2023blue}\\
         &  & electricity & -0.74 & MW & \\
         &  & $\mathrm{H_{2}}$ & -0.18 & tonne/hr & \\
         &  & nitrogen & -0.82 & tonne/hr & \\
         &  & cooling water & -155.14 & tonne/hr & \\
         \cline{2-6} 
         & steam methane\footnotemark[1] & \multicolumn{3}{l}{IHS} & \cite{IHS} \\
         & reforming (SMR) & \multicolumn{4}{l}{}\\
         \specialrule{.2em}{0em}{0em} 
         benzene\footnotemark[2] & methanol-to-aromatics & \multicolumn{3}{l}{IHS} & \cite{IHS}\\
        \cline{2-6}
        toluene\footnotemark[2] & solvent extraction from & \multicolumn{3}{l}{IHS} & \cite{IHS} \\
        xylene (mixed)\footnotemark[2] & pyrolysis gasoline& \multicolumn{4}{l}{}\\
         \specialrule{.2em}{0em}{0em} 
        carbon dioxide & point source capture & $\mathrm{CO_{2}}$ (100~bar) & 1 & tonne/hr & \cite{von2016selecting}\\
        ($\mathrm{CO_{2}}$) & (from processes with footnote\footnotemark[1]) & $\mathrm{CO_{2}}$ (1~bar) & -1 & tonne/hr & \\
        & & electricity & -0.1 & MW & \\
        &  & heat (high T)\footnotemark[3] & -0.003 & MW & \\
        \cline{2-6}
        & direct air capture\footnotemark[4] & $\mathrm{CO_{2}}$ (100~bar) & 1 & tonne/hr & \cite{deutz2021life}\\
        & & electricity (\textit{2016}) & -0.81 & MW & \\
        &  & heat (low T)\footnotemark[3] & -3.3 & MW & \\
        \cline{3-5}
        & & electricity (\textit{2020}) & -0.78 & MW & \\
        &  & heat (low T)\footnotemark[3] & -3.0 & MW & \\
        \cline{3-5}
        & & electricity (\textit{2025}) & -0.75 & MW & \\
        &  & heat (low T)\footnotemark[3] & -2.8 & MW & \\
        \cline{3-5}
        & & electricity (\textit{2030}) & -0.73 & MW & \\
        &  & heat (low T)\footnotemark[3] & -2.5 & MW & \\
        \cline{3-5}
        & & electricity (\textit{2035}) & -0.70 & MW & \\
        &  & heat (low T)\footnotemark[3] & -2.3 & MW & \\
        \cline{3-5}
        & & electricity (\textit{2040}) & -0.67 & MW & \\
        &  & heat (low T)\footnotemark[3] & -2.0 & MW & \\
        \cline{3-5}
        & & electricity (\textit{2045}) & -0.64 & MW & \\
        &  & heat (low T)\footnotemark[3] & -1.76 & MW & \\
         \specialrule{.2em}{0em}{0em} 
        carbon monoxide & reverse water-gas shift\footnotemark[1] & CO & 1 & tonne/hr & \cite{meys2021achieving}\\
         (CO) &  & water & 0.64 & tonne/hr & \\
         &  & $\mathrm{CO_{2}}$ (1 bar) & 0.075 & tonne/hr & \\
         &  & electricity & -0.3 & MW & \\
         &  & heat (high T)\footnotemark[3] & -0.6 & MW & \\
         &  & $\mathrm{H_{2}}$ & -0.08 & tonne/hr & \\
         &  & $\mathrm{CO_{2}}$ (100 bar) & -1.65 & tonne/hr & \\
         \specialrule{.2em}{0em}{0em} 
        cooling water & imports & cooling water &1 & tonne/hr&\\
         \specialrule{.2em}{0em}{0em} 
        ethylene\footnotemark[5]& methanol-to-olefins & \multicolumn{3}{l}{IHS} & \cite{IHS}\\
        propylene\footnotemark[5] & UOP/hydro& \multicolumn{4}{l}{}\\
        \cline{2-6}
        & steam cracking of naphtha & \multicolumn{3}{l}{IHS} & \cite{IHS}\\
        \specialrule{.2em}{0em}{0em} 
        heat (high T) & resistance heater & heat (high T)\footnotemark[3] & 1 & MW & \cite{meys2021achieving}\\
         &   & electricity & -1 & MW & \\
         &   & silicon carbide & -$9.3 \cdot 10^{-7}$ & tonne/hr & \\
         \cline{2-6}
         & $\mathrm{H_{2}}$ boiler & heat (high T)\footnotemark[3] & 1 & MW & \textit{own} \\
         &   & $\mathrm{H_{2}}$ & -0.03 & tonne/hr & \textit{calculation}\footnotemark[6]\\
        \specialrule{.2em}{0em}{0em} 
        hydrogen ($\mathrm{H_{2}}$) & PEM electrolysis & $\mathrm{H_{2}}$ & 1 & tonne/hr & \cite{Baumgartner.2021}\\
        &  & electricity (\textit{2016}) & -49.8 & MW & \\
        \cline{3-5}
        &  & electricity (\textit{2020}) & -48.3 & MW & \\
        \cline{3-5}
        &  & electricity (\textit{2025}) & -47.0 & MW & \\
        \cline{3-5}
        &  & electricity (\textit{2030}) & -45.7 & MW & \\
        \cline{3-5}
        &  & electricity (\textit{2035}) & -44.5 & MW & \\
        \cline{3-5}
        &  & electricity (\textit{2040}) & -43.4 & MW & \\
        \cline{3-5}
        &  & electricity (\textit{2045}) & -42.3 & MW & \\
         \cline{2-6}
        & steam methane\footnotemark[1] & \multicolumn{3}{l}{IHS} & \cite{IHS}\\
        & reforming (SMR) & \multicolumn{4}{l}{}\\
        \cline{2-6}
        & imports & $\mathrm{H_{2}}$ &1 & tonne/hr&\\
        \specialrule{.2em}{0em}{0em} 
        methane & carbon capture & methane &1 & tonne/hr&\cite{muller2011energiespeicherung,de2014parametric}\\
        & and utilization & electricity & -0.97 &MW &\\
        & & $\mathrm{H_{2}}$ & -0.51 & tonne/hr &\\
        & & $\mathrm{CO_{2}}$ & -2.78 &tonne/hr & \\
        \cline{2-6}
        & imports & methane &1 & tonne/hr&\\
        \specialrule{.2em}{0em}{0em} 
        methanol& carbon capture& methanol & 1 & tonne/hr & \cite{pereztechno}\\
        & and utilization & electricity & -0.018 & MW & \\
        & & heat (medium T)\footnotemark[3] & -0.44 & MW &\\
        & & $\mathrm{H_{2}}$ & -0.2 & tonne/hr&\\
        & & $\mathrm{CO_{2}}$& -1.46& tonne/hr&\\
        \cline{2-6}
        & from synthesis gas & \multicolumn{3}{l}{IHS} & \cite{IHS}\\
        \specialrule{.2em}{0em}{0em} 
        naphtha & imports & naphtha &1 & tonne/hr&\\
        \specialrule{.2em}{0em}{0em} 
        nitrogen&air separation by & \multicolumn{3}{l}{IHS} & \cite{IHS}\\
        & pressure-swing adsorption & \multicolumn{4}{l}{}\\
        \specialrule{.2em}{0em}{0em} 
        pyrolysis gas &steam cracking of naphtha & \multicolumn{3}{l}{IHS} & \cite{IHS}\\
        \specialrule{.2em}{0em}{0em} 
        steam & natural gas boiler & \multicolumn{3}{l}{IHS} & \cite{IHS}\\
        \cline{2-6}
        & electrode boiler & \multicolumn{3}{l}{IHS} & \cite{IHS}\\
        \specialrule{.2em}{0em}{0em} 
        silicon carbide & imports & silicon carbide &1 & tonne/hr&\\
        \specialrule{.2em}{0em}{0em} 
        synthesis gas (2:1) & steam methane reforming & \multicolumn{3}{l}{IHS} & \cite{IHS}\\
         &  with $\mathrm{H_{2}}$ skimming & \multicolumn{4}{l}{} \\
         \cline{2-6}
          & steam methane reforming & \multicolumn{3}{l}{IHS} & \cite{IHS}\\
         &  with $\mathrm{CO_{2}}$ import & \multicolumn{4}{l}{} \\
         \cline{2-6}
         &  partial oxidation of methane & \multicolumn{3}{l}{IHS} & \cite{IHS}\\
         \cline{2-6}
         &  mixing of CO and $\mathrm{H_{2}}$ & synthesis gas & 1 & tonne/hr & \cite{meys2021achieving}\\
         &  & CO & -0.875 & tonne/hr & \\
         &  & $\mathrm{H_{2}}$  & -0.125 & tonne/hr & \\
        \specialrule{.2em}{0em}{0em}   
\end{longtable}

\footnotetext[1]{Process produces concentrated $\mathrm{CO_{2}}$ at 1~bar for industrial point source capture.}
\footnotetext[2]{Aromatics process data was divided among the individual chemicals, benzene, toluene, and xylene, based on mass allocation.}
\footnotetext[3]{Temperature ranges for low, medium, and high temperature (T) heat are provided in \ref{tab:SI-temp_levels}.}
\footnotetext[4]{Process flows are linearly interpolated between the values given by Deutz and Bardow \cite{deutz2021life} for 'today', as 2016, and 'future', as 2050, as in Yang Shu et al. \cite{shu2023role}.}
\footnotetext[5]{Olefins process data was divided among the individual chemicals, ethylene and propylene, based on mass allocation.}
\footnotetext[6]{Assume 95\% efficiency from hydrogen LHV (120~MJ/kg $\mathrm{H_{2}}$) to MW heat}

\begin{table}[h]
\caption{Temperature levels for low, medium, and high temperature heat \cite{Baumgartner.2021}.}
\label{tab:SI-temp_levels}
\begin{center}
\begin{tabular}{ll}
\toprule
 & Temperature range [\SI{}{\degreeCelsius}]\\
\midrule
low temperature heat & $<100$ \\
medium temperature heat & 100 - 400 \\
high temperature heat & $>400$ \\
\bottomrule
\end{tabular}
\end{center}
\end{table}

\newpage
\renewcommand{\arraystretch}{1.2}
\small
\begin{longtable}{ll|llll|ll}
\captionsetup{justification=centering, width=\textwidth}
\caption{Cost and emission data for chemical industry model}
\label{tab:SI-costs_emissions}\\
\centering
         \textbf{Product} & \textbf{Process} & \multicolumn{4}{l|}{\textbf{Costs}} & \multicolumn{2}{l}{\textbf{Emissions}\footnotemark[1]} \\
\specialrule{.2em}{0em}{0em} 
ammonia & Haber-Bosch & CAPEX & 3300 & $\mathrm{\frac{k\euro}{tonne/hr}}$ & \cite{cesaro2021ammonia} & 0 &  \\
         &  & OPEX & 8.25\footnotemark[2] & $\mathrm{\frac{\euro}{tonne}}$ &  &  &  \\
\cline{2-8} 
         & steam methane & \multicolumn{3}{l}{IHS} & \multicolumn{3}{l}{\cite{IHS}}  \\
         & reforming (SMR) & \multicolumn{6}{l}{}\\
\specialrule{.2em}{0em}{0em} 
benzene\footnotemark[3] & MTA & \multicolumn{3}{l}{IHS} & \multicolumn{2}{l}{\cite{IHS}} &  \\
\cline{2-8}
toluene\footnotemark[3] & solvent extraction & \multicolumn{3}{l}{IHS} & \multicolumn{3}{l}{\cite{IHS}} \\
xylene (mixed)\footnotemark[3] & from pyrolysis gas & \multicolumn{6}{l}{}\\
\specialrule{.2em}{0em}{0em} 
carbon dioxide & point source capture & CAPEX & 119 & $\mathrm{\frac{k\euro}{tonne/hr}}$ & \cite{farla1995carbon} & 0 &  \\
         &  & OPEX & 0.43 & $\mathrm{\frac{\euro}{tonne}}$ &  &  &  \\
\cline{2-8} 
         & direct air capture\footnotemark[4] & CAPEX (\textit{2016/20}) & 5840 & $\mathrm{\frac{k\euro}{tonne/hr}}$ & \cite{fasihi2019techno} & -960 & \cite{deutz2021life} \\
         &  & OPEX (fixed)\footnotemark[5] & 4 & \% &  &  &  \\
\cline{3-8}
         &  & CAPEX (\textit{2025}) & 4272 &  &  & -966 &  \\
\cline{3-8}
         &  & CAPEX (\textit{2030}) & 2704 &  &  & -971 &  \\
\cline{3-8}
         &  & CAPEX (\textit{2035}) & 2300 &  &  & -976 &  \\
\cline{3-8}
         &  & CAPEX (\textit{2040}) & 1896 &  &  & -980 &  \\
\cline{3-8}
         &  & CAPEX (\textit{2045}) & 1744 &  &  & -984 &  \\
\specialrule{.2em}{0em}{0em} 
carbon monoxide & rWGS & \multicolumn{3}{l}{IHS} & \cite{IHS}\footnotemark[6] & 8 & \\
\specialrule{.2em}{0em}{0em} 
cooling water & imports & \multicolumn{3}{l}{IHS} & \cite{IHS} & ei 3.6  & \cite{ecoinvent3.6} \\
\specialrule{.2em}{0em}{0em} 
ethylene\footnotemark[3] & MTO & \multicolumn{3}{l}{IHS} & \multicolumn{3}{l}{\cite{IHS}}\\
propylene\footnotemark[3] & UOP/hydro & \multicolumn{6}{l}{} \\
\cline{2-8}
         & steam cracking & \multicolumn{3}{l}{IHS} & \multicolumn{3}{l}{\cite{IHS}} \\
         & of naphtha &  &  &  &  &  &  \\ 
\specialrule{.2em}{0em}{0em} 
heat (high T)\footnotemark[7] & resistance heater & CAPEX & 8,670 & $\mathrm{\frac{k\euro}{MW}}$ & \cite{geres2019roadmap} & $1 \cdot 10^{-3}$ & \\
         &  & OPEX & 2 & $\mathrm{\frac{\euro}{MWh}}$ &  &  &  \\
\cline{2-8}
         & $\mathrm{H_{2}}$ boiler & \multicolumn{6}{l}{\textit{assume same costs and emissions as resistance heater}} \\
\specialrule{.2em}{0em}{0em} 
hydrogen ($\mathrm{H_{2}}$) & PEM electrolysis & CAPEX (\textit{2016}) & 39,797 & $\mathrm{\frac{k\euro}{tonne/hr}}$ & \cite{Baumgartner.2021} & 0 & \\
         &  & OPEX (fixed) & 696 & $\mathrm{\frac{k\euro}{tonne/hr \cdot yr}}$ &  &  &  \\
\cline{3-6}
         &  & CAPEX (\textit{2020}) & 36,460 & $\mathrm{\frac{k\euro}{tonne/hr}}$ &  &  &  \\
         &  & OPEX (fixed) & 681 & $\mathrm{\frac{k\euro}{tonne/hr \cdot yr}}$ &  &  &  \\
\cline{3-6}
         &  & CAPEX (\textit{2025}) & 35,454 & $\mathrm{\frac{k\euro}{tonne/hr}}$ &  &  & \\
         &  & OPEX (fixed) & 615 & $\mathrm{\frac{k\euro}{tonne/hr \cdot yr}}$ &  &  & \\
\cline{3-6}
         &  & CAPEX (\textit{2030}) & 30,324 & $\mathrm{\frac{k\euro}{tonne/hr}}$ &  &  &  \\
         &  & OPEX (fixed) & 507 & $\mathrm{\frac{k\euro}{tonne/hr\cdot yr}}$ &  &  &  \\
\cline{3-6}
         &  & CAPEX (\textit{2035}) & 25,166 & $\mathrm{\frac{k\euro}{tonne/hr}}$ &  &  &  \\
         &  & OPEX (fixed) & 449 & $\mathrm{\frac{k\euro}{tonne/hr\cdot yr}}$ &  &  &  \\
\cline{3-6}
         &  & CAPEX (\textit{2040}) & 20,271 & $\mathrm{\frac{k\euro}{tonne/hr}}$ &  &  &  \\
         &  & OPEX (fixed) & 350 & $\mathrm{\frac{k\euro}{tonne/hr\cdot yr}}$ &  &  &  \\
\cline{3-6}
         &  & CAPEX (\textit{2045}) & 16,233 & $\mathrm{\frac{k\euro}{tonne/hr}}$ &  &  &  \\
         &  & OPEX (fixed) & 299 & $\mathrm{\frac{k\euro}{tonne/hr\cdot yr}}$ &  &  &  \\
\cline{2-8}
         & steam methane & \multicolumn{3}{l}{IHS} & \multicolumn{3}{l}{\cite{IHS}} \\
         & reforming (SMR) & \multicolumn{6}{l}{}\\
\cline{2-8}
         & imports & OPEX\footnotemark[8] & $1 \cdot 10^{5}$ & $\mathrm{\frac{k\euro}{tonne}}$ &  & ei 3.5 & \cite{wernet2016ecoinvent} \\
\specialrule{.2em}{0em}{0em} 
methane & carbon capture & CAPEX (\textit{2016/20}) & 11,536 & $\mathrm{\frac{k\euro}{tonne/hr}}$ & \cite{energiewende2018future} & 38 &  \\
         & and utilization & OPEX (fixed)\footnotemark[5] & 3 & \% &  &  &  \\
\cline{3-6}
         &  & CAPEX (\textit{2025}) & 10,811 & $\mathrm{\frac{k\euro}{tonne/hr}}$ &  &  & \\
\cline{3-6}
         &  & CAPEX (\textit{2030}) & 10,086 & $\mathrm{\frac{k\euro}{tonne/hr}}$ &  &  & \\
\cline{3-6}
         &  & CAPEX (\textit{2035}) & 9,493 & $\mathrm{\frac{k\euro}{tonne/hr}}$ &  &  & \\
\cline{3-6}
         &  & CAPEX (\textit{2040}) & 8,899 & $\mathrm{\frac{k\euro}{tonne/hr}}$ &  &  & \\
\cline{3-6}
         &  & CAPEX (\textit{2045}) & 8,305 & $\mathrm{\frac{k\euro}{tonne/hr}}$ &  &  & \\
\cline{2-8}
         & imports & OPEX & 0.5 & $\mathrm{\frac{k\euro}{tonne}}$ & \cite{Baumgartner.2021} & ei 3.5 & \cite{wernet2016ecoinvent} \\
\specialrule{.2em}{0em}{0em} 
methanol\footnotemark[3] & carbon capture & \multicolumn{3}{l}{IHS} & \cite{IHS} & 1460 &  \\
         & and utilization &  &  &  &  &  &  \\
\cline{2-8}
         & from synthesis gas & \multicolumn{3}{l}{IHS} & \multicolumn{3}{l}{\cite{IHS}}  \\
\specialrule{.2em}{0em}{0em} 
naphtha & imports & OPEX & 0.44 & $\mathrm{\frac{k\euro}{tonne}}$ & \cite{iea2018} & ei 3.6 & \cite{ecoinvent3.6} \\
\specialrule{.2em}{0em}{0em} 
nitrogen & air separation by & \multicolumn{3}{l}{IHS} & \multicolumn{3}{l}{\cite{IHS}} \\
         & pressure-swing & \multicolumn{6}{l}{}\\
         & adsorption &  &  &  &  &  &  \\ 
\specialrule{.2em}{0em}{0em} 
pyrolysis gas & steam cracking & \multicolumn{3}{l}{IHS} & \multicolumn{3}{l}{\cite{IHS}} \\
         & of naphtha &  &  &  &  &  &  \\ 
\specialrule{.2em}{0em}{0em} 
steam & natural gas boiler & \multicolumn{3}{l}{IHS} & \multicolumn{3}{l}{\cite{IHS}}  \\
\cline{2-8}
         & electrode boiler & \multicolumn{3}{l}{IHS} & \multicolumn{3}{l}{\cite{IHS}} \\
\specialrule{.2em}{0em}{0em} 
silicon carbide & imports & \multicolumn{3}{l}{IHS} & \cite{IHS} & ei 3.6 & \cite{ecoinvent3.6} \\
\specialrule{.2em}{0em}{0em} 
synthesis gas (2:1) & steam methane & \multicolumn{3}{l}{IHS} & \multicolumn{3}{l}{\cite{IHS}} \\
         & reforming with & \multicolumn{6}{l}{}\\
         & $\mathrm{H_{2}}$ skimming & \multicolumn{6}{l}{}\\
\cline{2-8}
         & steam methane & \multicolumn{3}{l}{IHS} & \multicolumn{3}{l}{\cite{IHS}} \\
         & reforming with & \multicolumn{6}{l}{}\\
         & $\mathrm{CO_{2}}$ import & \multicolumn{6}{l}{}\\
\cline{2-8}
         & partial oxidation & \multicolumn{3}{l}{IHS} & \multicolumn{3}{l}{\cite{IHS}} \\
         & of methane &  &  &  &  &  &  \\ 
\cline{2-8}
         & mixing of & CAPEX & 0 & $\mathrm{\frac{k\euro}{tonne/hr}}$ &  & 1.2 &  \\
         & CO and $\mathrm{H_{2}}$\footnotemark[9] & OPEX & 0 & $\mathrm{\frac{\euro}{tonne}}$ &  &  &  \\
\specialrule{.2em}{0em}{0em} 
\end{longtable}

\footnotetext[1]{Refers to process operational emissions in $\mathrm{[\frac{kg\,CO_{2}\text{-}eq}{process\,functional\,unit}]}$. Life cycle operational emissions of process inputs that are modelled separately (i.e. $\mathrm{H_{2}}$) are also considered separately. Except for direct air capture, which includes the life cycle impacts of the adsorbent, process operational emissions are calculated by closing the atom balance around process flows, as done by Meys et al.\cite{meys2021achieving}.}
\footnotetext[2]{Assume 2\% O\&M factor and 8000 full-load hours.}
\footnotetext[3]{Emissions include end-of-life emissions assuming complete combustion into $\mathrm{CO_{2}}$, as done by Zibunas et al.\cite{zibunas2022cost}.}
\footnotetext[4]{Intermediate years 2025, 2035, and 2045 are linearly interpolated using the data from Fasihi et al. \cite{fasihi2019techno}.}
\footnotetext[5]{\% for OPEX calculation refers to \% of CAPEX per year.}
\footnotetext[6]{Own calculation derived from conventional CO production \cite{IHS}. Take ratio of mass output per mass CO for conventional process, $\mathrm{r_{conv}}$, and for reverse water-gas shift (rWGS) process, $\mathrm{r_{rWGS}}$, and calculate scaling factor as $\mathrm{(\frac{r_{rWGS}}{r_{conv}})^{0.6}}$. Scaling factor is used on conventional process costs to derive costs for rWGS process.}
\footnotetext[7]{CAPEX taken from \cite{geres2019roadmap}, as investment cost for electric cracker. OPEX assumes 0.2\% of the yearly CAPEX as the largest OPEX cost contributor is electricity which is accounted for separately.}
\footnotetext[8]{We place a high price penalty on imported green $\mathrm{H_{2}}$ such that the system prioritizes domestic energy resources.}
\footnotetext[9]{Mixing cost neglected, as by Zibunas et al.\cite{zibunas2022cost}.}

\newpage
\section{Energy system model modifications}
\label{sec:SI-ESOM_modifications}
In this section, we provide the modifications made to the original German energy system model from Baumgärtner et al. \cite{Baumgartner.2021}. \ref{tab:SI-HT_heat_demands} provides modifications to the annual exogenous high temperature heat demand, which is distributed evenly for every hour. Of the exogenous heat demands, changes were made only for high temperature heat because chemical process heat requirements for conventional production were modelled as high temperature. \ref{tab:SI-electricity_demands} provides modifications to the annual exogenous electricity demand. The chemical industry electricity demand is assumed constant for every hour and is subtracted from the original hourly profile from \cite{Baumgartner.2021}. \ref{tab:SI-emissions} provides modifications to the exogenous emissions targets. 

\begin{table}[H]
\centering
\caption{Chemical industry high temperature (T) heat demands subtracted from the exogenous demands from Baumgärtner et al.\cite{Baumgartner.2021}.}
\label{tab:SI-HT_heat_demands}
\begin{tabular}{ll}
\toprule
 & High T Heat [TWh/yr]\\
\midrule
\textbf{Baumgärtner et al.\cite{Baumgartner.2021}} & \textbf{273} \\
Chemical Processing \cite{Baumgartner.2021}
 & -59 \\
 Net exogenous demand & 214\\
\bottomrule
\end{tabular}
\end{table}

\begin{table}[H]
\caption{Chemical industry electricity demands subtracted from the exogenous demands from Baumgärtner et al.\cite{Baumgartner.2021}.}
\label{tab:SI-electricity_demands}
\begin{center}
\begin{tabular}{ll}
\toprule
 & Electricity [TWh/yr]\\
\midrule
\textbf{Baumgärtner et al.\cite{Baumgartner.2021}} & \textbf{550} \\
Mechanical energy \cite{repenning2015klimaschutzszenario}
 & -35.6 \\
Process cooling \cite{repenning2015klimaschutzszenario}
 & -1.8 \\
 \bottomrule
 Net exogenous demand & 512.6\\
 \hline
\end{tabular}
\end{center}
\end{table}

\begin{table}[H]
\caption{Modified exogenous emission targets in $\mathrm{Mt_{CO_{2}\text{-}eq}}$/yr.}
\label{tab:SI-emissions}
\begin{center}
\begin{tabular}{llllllll}
\toprule
 & 2016&2020&2025&2030&2035&2040&2045\\
\midrule
\textbf{Baumgärtner et al.\cite{Baumgartner.2021}} & \textbf{690}&\textbf{617}&\textbf{526}&\textbf{435}&\textbf{362}&\textbf{290}&\textbf{217} \\
Updated reduction targets\footnotemark[1] \cite{bundestag2021erstes} & 700 &571&441&337&219&102&0 \\
Process emissions \cite{FederalMinistryfortheEnvironmentNatureConservationandNuclearSafety.} & +7 &+7&+7&+4&+4&+4&+0 \\
Chemical use-phase emissions\footnotemark[2] & +35 &+35&+35&+12&+12&+12&+0 \\
 \hline
 Net emission targets & 742 &613&483&354&236&118&0 \\
\bottomrule
\end{tabular}
\end{center}
\end{table}

\footnotetext[1]{Historical emissions were updated using the BMU 2021 report \cite{FederalMinistryfortheEnvironmentNatureConservationandNuclearSafety.} rather than BMU 2018 used in \cite{Baumgartner.2021}.}
\footnotetext[2]{Use-phase emissions calculated based on complete combustion of the carbon-containing chemicals to $\mathrm{CO_{2}}$, with a 65\% reduction by 2030 and net-zero by 2045.}

\section{Underlying supply chain modelling for the \textit{Cost-Avoided}}
\label{sec:SI-supply_chain_modelling}
\begin{figure}[H]    
\includegraphics[width=0.7\textwidth]{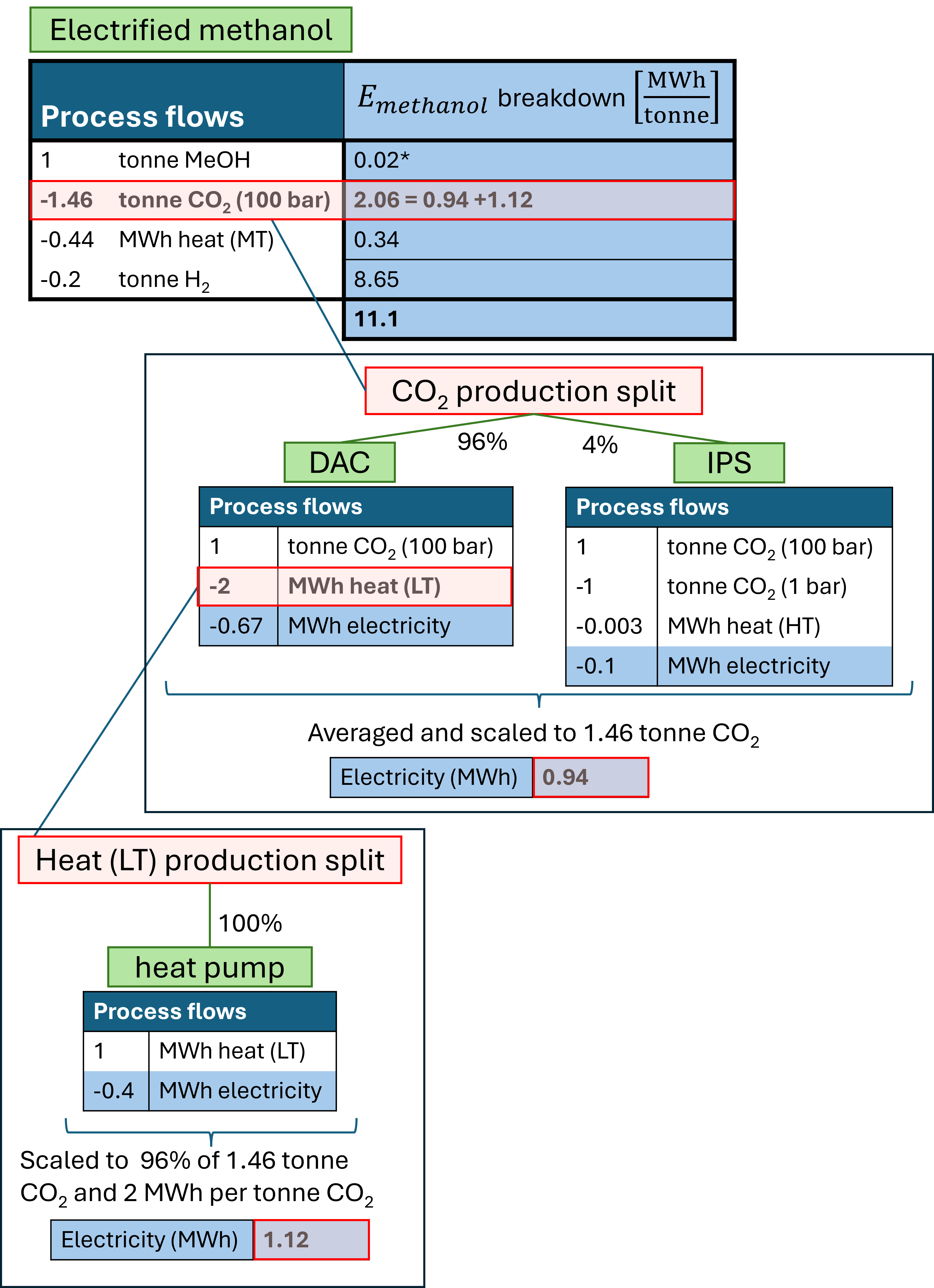}
    \centering
    \caption{Depiction of underlying $\mathrm{CO_{2}}$ supply chain consideration in the calculation of methanol's electrified production electricity demand, $E_{methanol}$. Other process flows follow the same procedure for calculation of their contribution to $E_{methanol}$. The operating costs ($C_{methanol}^{op,elec}$, $C_{methanol}^{op,fossil}$) and emissions ($e_{methanol}^{CO_{2},elec}$, $e_{methanol}^{CO_{2},fossil}$) are also calculated in the same manner. The process flows are taken from the year 2040, and the production mix as the aggregate production mix from 2040. \textbf{TOP}: Material and energy process flows that go into production of 1~tonne electrified methanol, and the electricity demands associated with a given flow. \textbf{MIDDLE}: Zoom into $\mathrm{CO_{2}}$ production mix between direct air capture (DAC) and industrial point source (IPS), showing how the electricity requirements are averaged across the production mix and scaled to the methanol feed requirement. \textbf{BOTTOM}: Zoom into low temperature (LT) heat production mix, showing how the electricity demands are scaled to the methanol $\mathrm{CO_{2}}$ feed requirement and to the portion produced via DAC. We note that while here we show a static example of the underlying supply chain, the supply chain varies for every hour. *direct electricity demand of electrified methanol production.}
    \label{fig:SI-background_system}
\end{figure}

To calculate the electricity demands ($E_{i,t}$), operating costs ($C_{i,t}^{op,elec}$, $C_{i,t}^{op,fossil}$) and emissions ($e_{i,t}^{CO_{2},elec}$, $e_{i,t}^{CO_{2},fossil}$) for the \textit{Cost-Avoided}, we consider both the direct process and the underlying supply chains of the process material and energy inputs (\ref{fig:SI-background_system}). In considering the underlying supply chains, we assume 100\% allocation to a process's reference product. For example, $\mathrm{H_{2}}$ as a by-product from synthesis gas production would not contribute to the electricity, operating costs, and emissions of an electrified process that utilizes that $\mathrm{H_{2}}$ because everything is allocated to the synthesis gas. 

Point-source $\mathrm{CO_{2}}$ emissions are released from fossil-based processes (i.e. ammonia and hydrogen via steam-methane reforming) assuming a pressure of 1~bar (Table~\ref{tab:SI-flows}). These emissions can then be compressed to 100~bar for use as feedstock in CCU-based electrified chemical processes. We allocate these point-source emissions to the CCU-based processes rather than to the fossil-based processes since the CCU-based processes should account for the underlying emissions of the $\mathrm{CO_{2}}$ feed \cite{muller2020guideline}. Our allocation assumption, however, underestimates the emissions from fossil-based processes when the point-source emissions are not utilized downstream, consequently decreasing the \textit{Cost-Avoided} from electrified production under these circumstances. This phenomenon can be seen in the ammonia portion of the load-duration curve presented in the main paper (Figure~5, MIDDLE). Due to our allocation assumption, the \textit{Cost-Avoided} from electrified ammonia production is underestimated for the portion of fossil-based ammonia that produces point-source $\mathrm{CO_{2}}$ in excess of what is needed to meet the CCU-based methanol demand. As a result, there is a tranche of ammonia production that is prioritized over high-temperature heat electrified production but that is shown to have a lower \textit{Cost-Avoided} in the merit order curve (Figure~5, BOTTOM, dashed red line). This phenomenon can also be seen in the results for our sensitivity on the HT heat cost assumptions (Supplementary Section~\ref{sec:SI-scenario_low_capex}). This scenario electrifies only methanol and the portion of ammonia in excess of the fossil-based ammonia that produces the necessary point-source $\mathrm{CO_{2}}$ to meet the electrified methanol demand (\ref{fig:SI-flexibility_provision}).

In addition to our allocation assumption regarding point-source emissions, we consider 100~bar $\mathrm{CO_{2}}$ flows into a process as negative emissions, and out of a process as positive emissions. We also account for a process's  modelled $\mathrm{CO_{2}}$-eq operational emissions to properly account for the carbon flows in the underlying supply chain of an electrified process.  

In \ref{fig:SI-background_system} we show an example of how the underlying $\mathrm{CO_{2}}$ supply chain is considered in the electricity demand of methanol ($E_{methanol}$). This procedure applies to all the process flows, and to all the flows in the processes of the underlying supply chain. The top table shows the material and energy flows involved in the production of 1~tonne methanol, and their corresponding contributions to the electricity demand of methanol of $\mathrm{11.1~\frac{MWh}{tonne~methanol}}$. The direct process electricity demand is listed next to the 1~tonne of methanol. In the middle part of the figure, we zoom into the production mix of $\mathrm{CO_{2}}$, showing that 96\% is produced via direct air capture (DAC) and 4\% from industrial point sources. We calculate a weighted average of the electricity demands using the production split, and scale this weighted electricity demand by $\mathrm{1.46~\frac{tonne~CO_{2}}{tonne~methanol}}$ to obtain $\mathrm{0.94~\frac{MWh}{tonne~methanol}}$ attributed to $\mathrm{CO_{2}}$ production. In the bottom part of the figure, we zoom into the production of low temperature (LT) heat needed for DAC, which is 100\% produced via heat pumps. We take the electricity demand per MWh of LT heat, and scale it to 96\% of the $\mathrm{1.46~\frac{tonne~CO_{2}}{tonne~methanol}}$ multiplied by $\mathrm{2~\frac{MWh~LT~heat}{tonne~CO_{2}}}$ to obtain $\mathrm{1.12~\frac{MWh}{tonne~methanol}}$ attributed to heat production for the $\mathrm{CO_{2}}$ supply chain. With this example, we show how the $\mathrm{2.06~\frac{MWh}{tonne~methanol}}$ attributed to the $\mathrm{CO_{2}}$ process flow is calculated.

\section{Sector-specific transition pathway results}
\label{sec:SI-detailed_transitions}
\begin{figure}[h]
\centering
\includegraphics[width=0.8\textwidth]{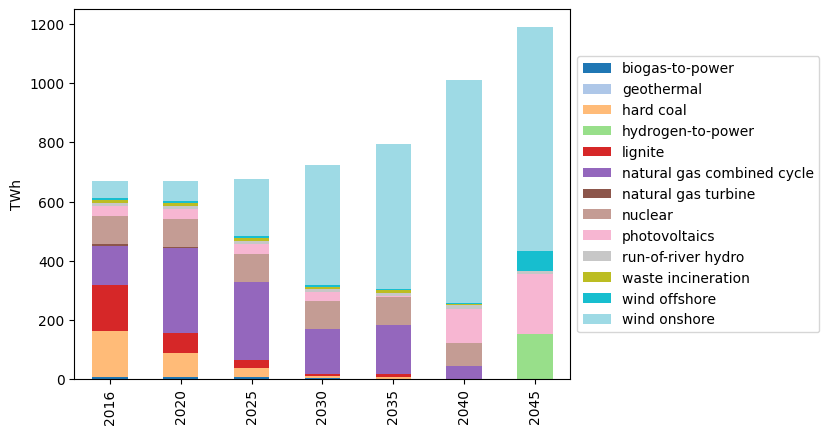}
    \caption{Electricity production mix transition pathway. A nuclear phase-out is exogenously imposed by 2045, as shown by the resulting production mix in that year.}
    \label{fig:SI-elec_transition}
\end{figure}

\begin{figure}[h]
\centering
\includegraphics[width=0.8\textwidth]{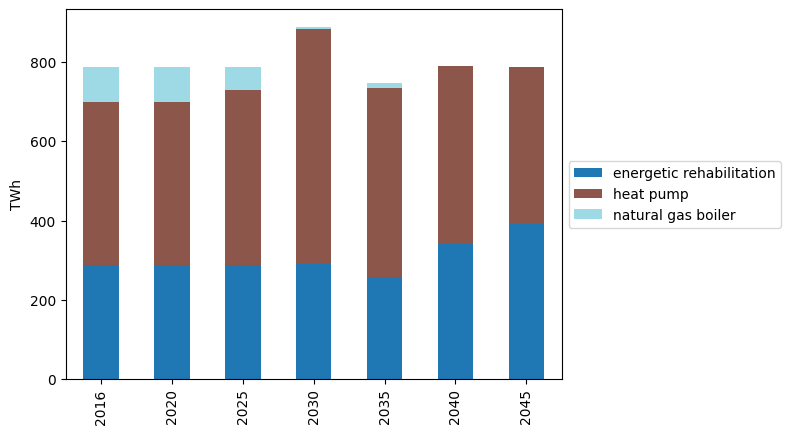}
    \caption{Residential heat production mix transition pathway. Varying aggregate production arises from time series aggregation. Energetic rehabilitation refers to building renovations with improved thermal insulation. We introduce an exogenous constraint limiting the heat provision from energetic rehabilitation to a maximum of 50\%, as in Baumgärtner et al. \cite{Baumgartner.2021}.}
    \label{fig:SI-civil_transition}
\end{figure}

\begin{figure}[h]
\centering
\includegraphics[width=0.8\textwidth]{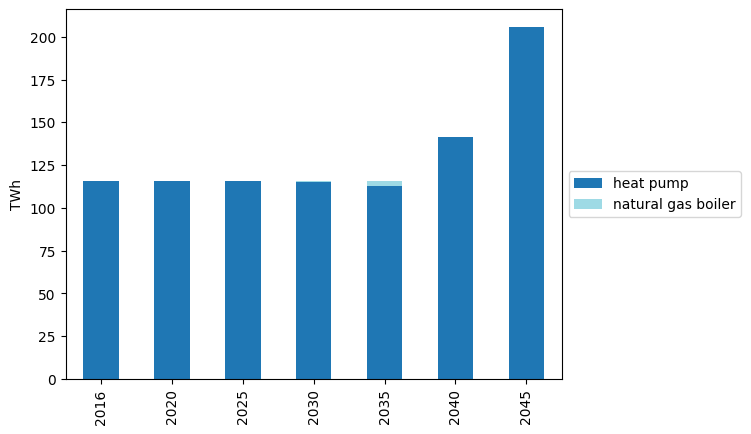}
    \caption{Low temperature (LT) industrial heat production mix transition pathway. Our cost-optimal results how that heat pumps should comprise 100\% of LT heat production since 2016. Natural gas boilers are introduced in 2035 due to the increased electricity demand from the mobility fleet transition to electric vehicles, which reduces the renewable electricity available for heat pumps.}
    \label{fig:SI-low_transition}
\end{figure}

\begin{figure}[h]
\centering
\includegraphics[width=0.8\textwidth]{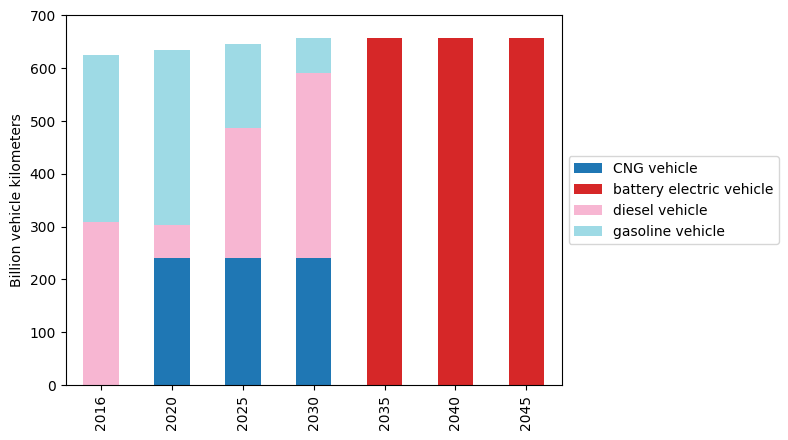}
    \caption{Mobility production mix transition pathway. The transition of the mobility sector occurs in a single investment period in 2035 due to the existing vehicle fleet reaching the end of its lifetime, and the foresight to the net-zero emissions target in 2045 influencing the decision to invest in electric vehicles rather than reinvest in fossil fuel vehicles. CNG: compressed natural gas.}
    \label{fig:SI-mobility_transition}
\end{figure}

\begin{figure}[h]
\centering
\includegraphics[width=0.8\textwidth]{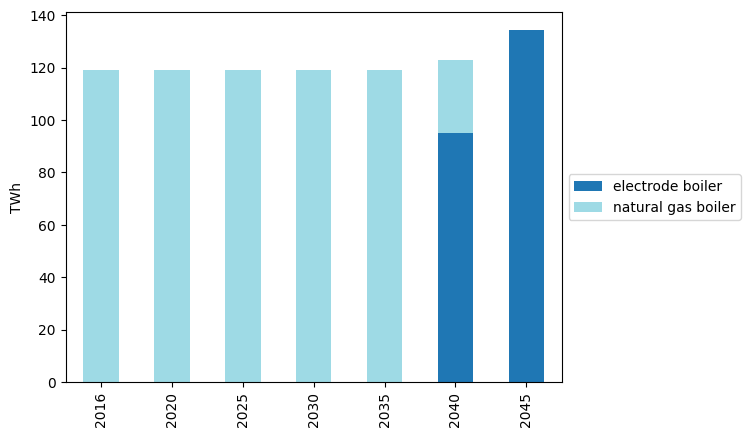}
    \caption{Medium temperature (MT) industrial heat production mix transition pathway. An increase in methanol production as an intermediate for electrified olefins and aromatics leads to an increase in MT heat production in 2040 and 2045.}
    \label{fig:SI-medium_transition}
\end{figure}

\begin{figure}[h]
\centering
\includegraphics[width=0.8\textwidth]{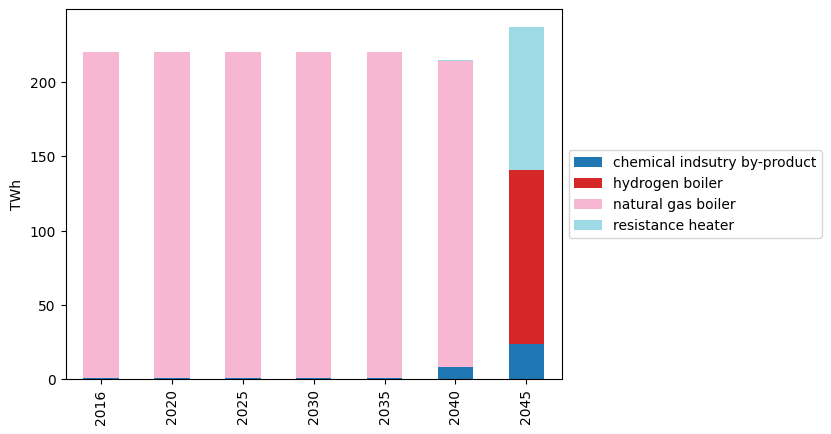}
    \caption{High temperature (HT) industrial heat production mix transition pathway. Only 0.5\% of HT heat is produced via resistance heaters in 2040. In 2045, all HT heat transitions away from natural gas boilers due to the exogenous net-zero emissions constraint. The late transition of high temperature heat to electrified technologies is driven by the high capital cost assumptions for resistance heaters and hydrogen boilers. Results considering optimistic cost assumptions are presented in Supplementary Section~\ref{sec:SI-scenario_low_capex}}.
    \label{fig:SI-high_transition}
\end{figure}

In this section, we present the transition pathwas results for the individual energy sectors: electricity (\ref{fig:SI-elec_transition}), residential heat (\ref{fig:SI-civil_transition}), low (\ref{fig:SI-low_transition}), medium (\ref{fig:SI-medium_transition}), and high (\ref{fig:SI-high_transition}) temperature industrial heat, and mobility (\ref{fig:SI-mobility_transition}).

\clearpage
\section{Sensitivity on high-temperature heat cost assumptions}
\label{sec:SI-scenario_low_capex}
Our base scenario presented in the main paper assumes high costs for high-temperature (HT) resistance heaters and $\mathrm{H_{2}}$ boilers (\ref{tab:SI-costs_emissions}) due to the low TRL for large-scale industrial deployment. Here, we present the results for a scenario with optimistic cost assumptions for HT heat technologies (\ref{tab:SI-HT_costs}). For HT resistance heaters, we assume the same costs as for medium-temperature heat electrode boilers. For $\mathrm{H_{2}}$ boilers, we assume the same costs as for natural gas boilers. The cost values are taken from Baumgärtner et al. \cite{Baumgartner.2021}.

\begin{figure}[h]
\centering
\includegraphics[width=0.8\textwidth]{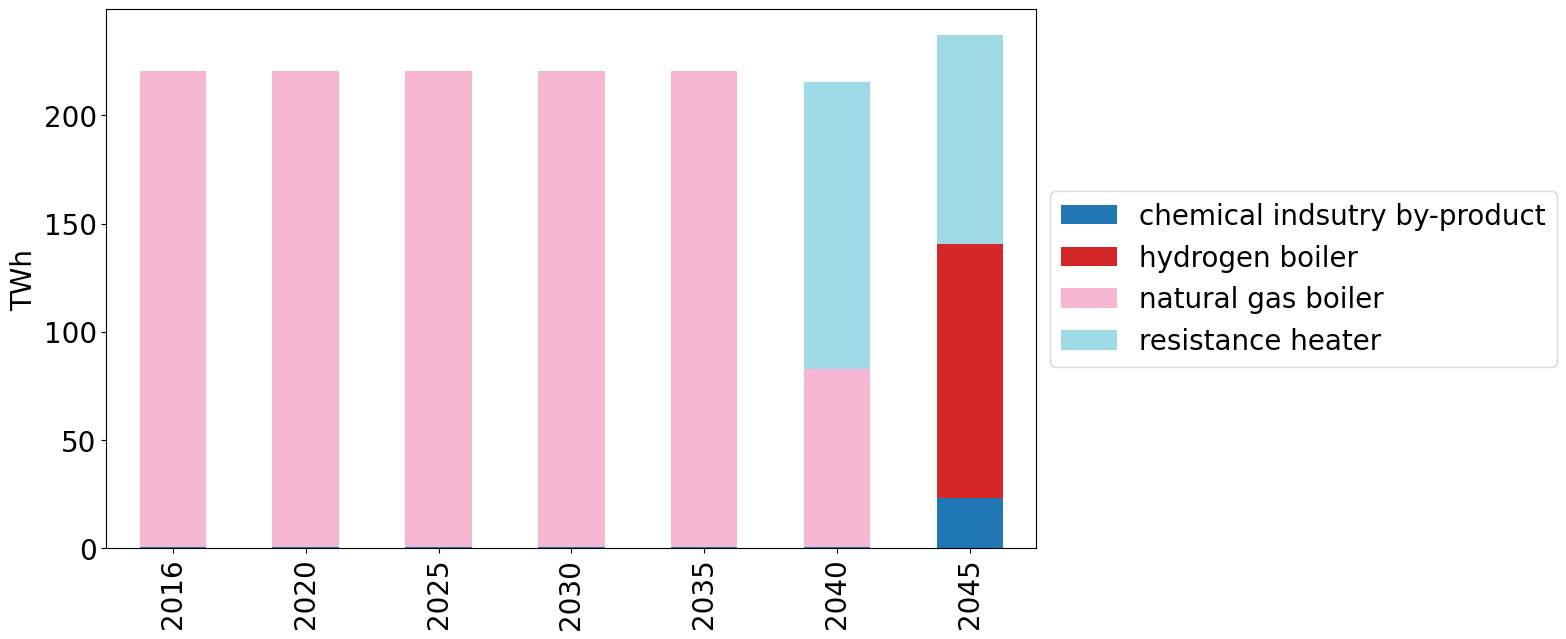}
    \caption{High temperature (HT) industrial heat production mix transition pathway under optimistic cost assumptions for resistance heaters and $\mathrm{H_{2}}$ boilers. 62\% of HT heat is produced via resistance heaters in 2040, as opposed to only 0.5\% in the base scenario.}
    \label{fig:SI-high_transition_2}
\end{figure}

\begin{figure}[h]
\centering
\includegraphics[width=0.6\textwidth]{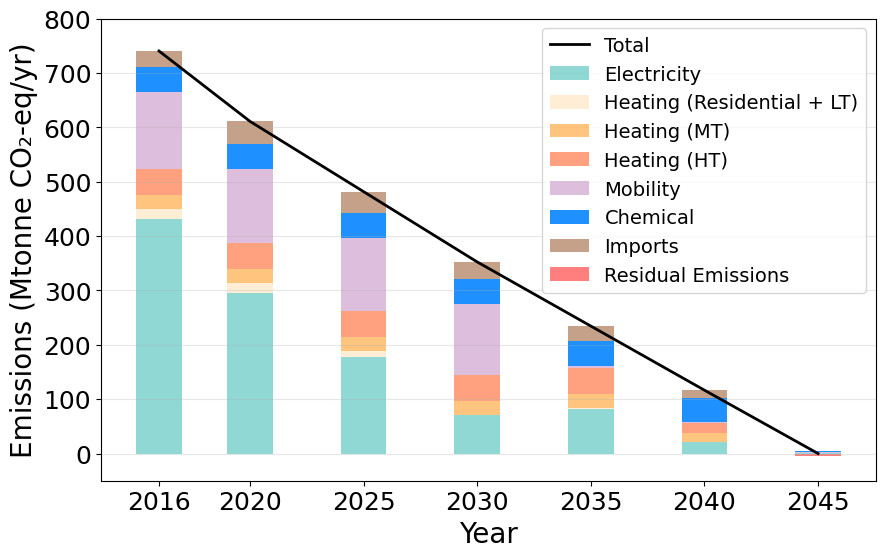}
    \caption{$\mathrm{CO_{2}}$-eq emissions, in million tonnes, of the integrated energy system and chemical industry model by sector. With the optimistic cost assumptions, high-temperature (HT)  heat transitions in 2040, rather than 2045, pushing the majority of the chemical industry to transition last in 2045. MT: medium temperature, LT: low temperature.} 
 \label{fig:SI-transition_pathway}
\end{figure}

\begin{figure}[h]
\centering
\includegraphics[width=0.6\textwidth]{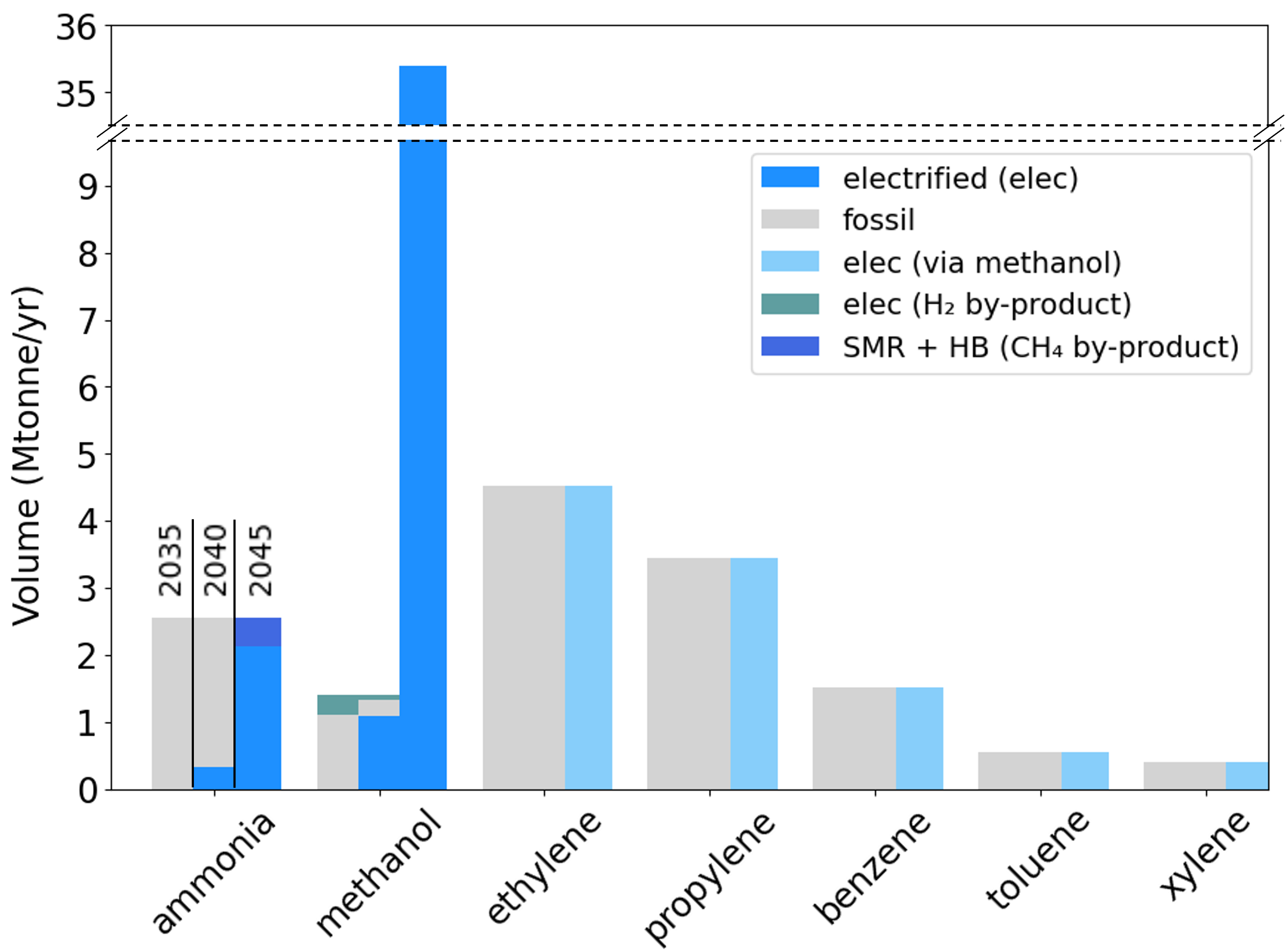}
     \caption{Transition of chemical production from 2035 to 2045. Years prior to 2035 have the same production mix as 2035 and are therefore excluded from the figure. With the optimistic cost assumptions, chemicals still begin their transition to electrified production in 2040. However, only methanol and some ammonia are produced electrically in 2040, while the olefins and the bulk of ammonia transition in 2045. elec (via methanol) refers to the methanol-to-olefins and methanol-to-aromatics processes. elec ($\mathrm{H_{2}}$ by-product) refers to methanol produced via  CCU using by-product $\mathrm{H_{2}}$ from synthesis gas production and $\mathrm{CO_{2}}$ from chemical industry point sources. SMR + HB ($\mathrm{CH_{4}}$ by-product) refers to ammonia produced via  steam methane reforming + Haber-Bosch using by-product $\mathrm{CH_{4}}$ from other electrified processes.  } 
 \label{fig:SI-chem_ind_transition}
\end{figure}

Our results show that under optimistic cost assumptions for HT heat technologies, 62\% of HT heat is produced via resistance heaters in 2040 as opposed to only 0.5\% in the base scenario (\ref{fig:SI-high_transition_2}). The energy system invests in resistance heater capacity to meet 100\% of the HT heat demand and in electrified chemical production capacities for the portion of the chemical industry that has a higher \textit{Cost-Avoided} than the HT heat sector. This portion includes 100\% of the hourly methanol demand and 16\% of the hourly ammonia demand. Electrification of the remaining ammonia demand and the olefins, which are electrified in the base scenario, is pushed to 2045. The 16\% electrified ammonia portion corresponds to the difference between the hourly demand and the fossil-based production that produces the necessary point-source $\mathrm{CO_{2}}$ to meet the electrified, CCU-based, methanol demand. The \textit{Cost-Avoided} for the share of electrified ammonia is higher than for the remaining ammonia portion because the fossil-based process would release point-source $\mathrm{CO_{2}}$ emissions that cannot be further utilized by CCU-based methanol (Supplementary Section~\ref{sec:SI-supply_chain_modelling}). Thus, the electrified process abates these emissions, increasing the \textit{Cost-Avoided} and making investments in this portion of electrified ammonia beneficial. Still, even though the electrified olefins and a portion of electrified ammonia production are pushed to 2045 under this optimistic cost scenario (\ref{fig:SI-transition_pathway}), methanol and ammonia begin their transitions in 2040 (\ref{fig:SI-chem_ind_transition}). This finding emphasizes the alignment between the transitions of the chemical industry and the HT heat sector, and highlights the prioritization of methanol and ammonia among chemicals for electrification.

\begin{figure}[h]
\centering
\includegraphics[width=0.9\textwidth]{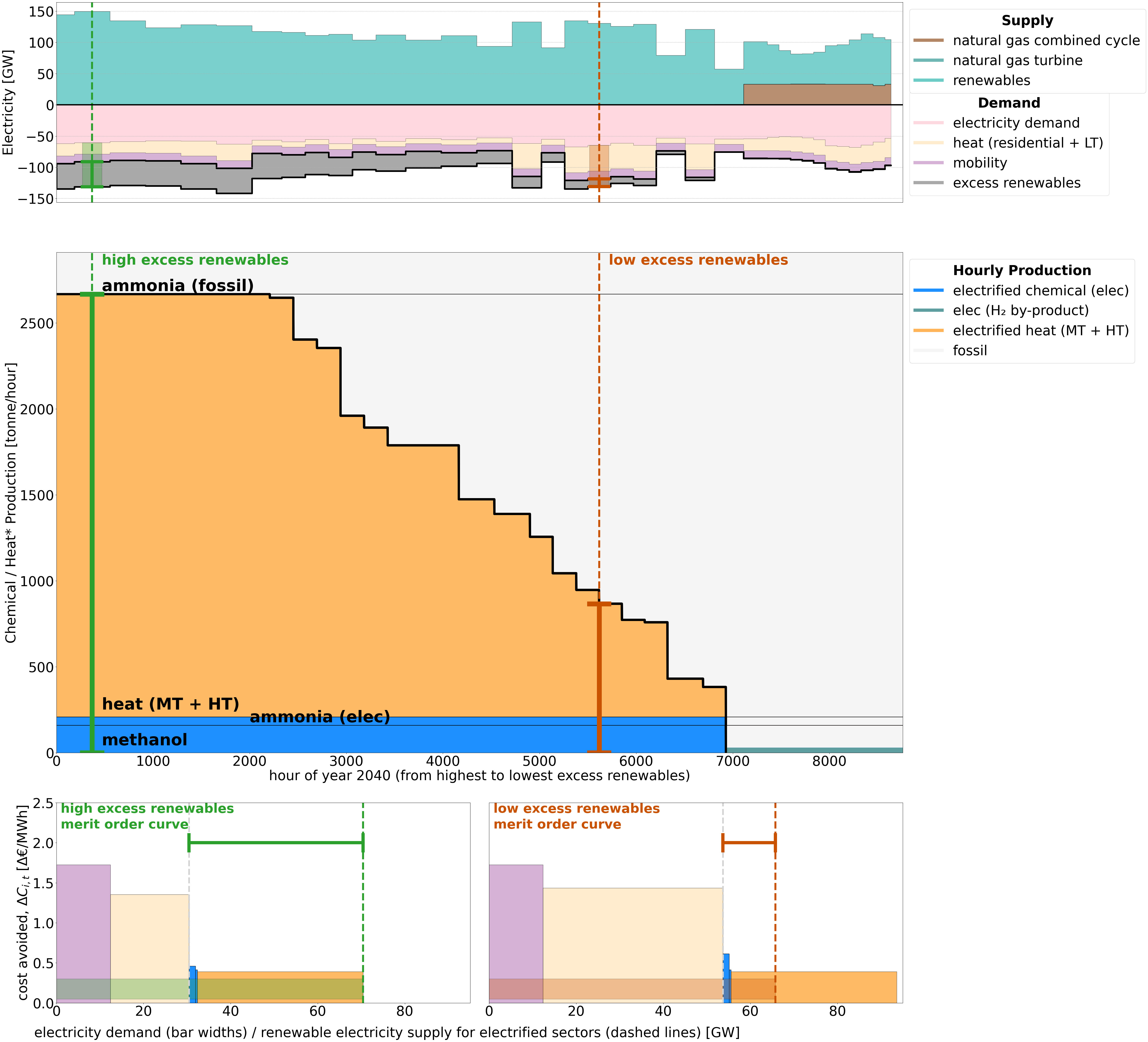}
    \caption{\textbf{TOP}: electricity supply and demand for each hour in the year 2040. The hours are ordered from highest to lowest excess renewables after full electrification of the electricity, residential and low-temperature (LT) heat, and mobility sectors. Due to the time series aggregation, the hours repeat themselves, causing the steps in the figure.  Electricity storage is excluded from the figure. With the optimistic cost assumptions, electricity consumption for high-temperature heat increases, decreasing the availability for electrified chemicals. \textbf{MIDDLE}: load-duration curve for the year 2040, with the hourly breakdown of electrified vs. fossil-based production for medium and high-temperature (MT + HT) heat and for chemicals with electrified capacities. The hours are in the same order as in the electricity balance plot (top figure). Ammonia is split into its electrified (elec) and fossil-based portions due to the different \textit{Cost-Avoided} for each portion (Supplementary Section~\ref{sec:SI-scenario_low_capex}). In hours with high excess renewables (dashed green line), all chemicals and heat are produced via their electrified process up to the installed capacities. In hours with low excess renewables (dashed red line), only a portion of heat is electrified while all chemical electrified capacities are operated. This behavior is explained by the merit order curves created by the \textit{Cost-Avoided} and the electricity demand of each electrified product (bottom figure). \textbf{BOTTOM}:  merit order curves of electrified products in the sector-coupled energy system. Each curve corresponds to a separate hour, identified by the red and green dashed lines crossing the top and middle figures. The red and green dashed lines show the hourly renewable electricity supplied for the electrified energy sectors. Everything to the left of the intersection between an hour’s renewable electricity supply and the merit order curve is produced electrically for that hour. Mobility, residential heat, and LT heat are fully electrified in every hour (dashed gray lines). The excess renewables (red and green brackets) are then used for electrification of chemicals and MT + HT heat. *Heat production is shown in tonne natural gas equivalents using a heating value of $\mathrm{15.4~\frac{MWh}{tonne}}$.} 

 \label{fig:SI-flexibility_provision}
\end{figure}

Under optimistic cost assumptions, we see diminished flexibility provision from the chemical industry to the energy system in 2040 and increased flexibility from the MT and HT heat sectors (\ref{fig:SI-flexibility_provision}). In 80\% of the hours in the year, all of the electrified chemicals installed capacities are used, showing less benefit from diversifying and over-sizing the chemical industry than in our base results. However, it must be noted that due to our system set-up, we are not able to capture any flexibility provision from a transitioning chemical industry beyond 2040 since we force the chemical industry to fully electrify by 2045. In actuality, the chemical industry will likely not fully electrify but will rather be comprised of a mix of sustainable production options together with electrification, such as biomass, recycling, and carbon capture and storage. Once this mix is available, it can enable flexibility through diversification and over-capacities regardless of when the chemical industry's transition occurs relative to the other energy sectors. Furthermore, our results indicate that flexibility can also arise from other energy sectors that integrate electrified technologies with more dispatchable alternatives.

In summary, our scenario with optimistic cost assumptions for HT resistance heaters and $\mathrm{H_{2}}$ boilers shows that the HT heat sector transitions before the bulk of the chemical industry. Still, methanol and ammonia begin their transitions alongside the HT heat sector, highlighting their priority for electrification within the chemical industry. Finally, although the flexibility provision from a transitioning chemical industry is weakened under this optimistic cost scenario, we see increased flexibility from a transitioning HT heat sector. This finding shows the broader benefit of over-sizing and diversifying production options across the energy system.

\begin{table}[H]
\caption{High-temperature heat technology cost assumptions for optimistic scenario.}
\label{tab:SI-HT_costs}
\begin{center}
\begin{tabular}{llll}
\toprule
 Process & CAPEX [$\mathrm{\frac{k\euro}{MW}}$] & OPEX (fixed) [$\mathrm{\frac{k\euro}{MW \cdot yr}}$]&\\
\midrule
resistance heater & 238 & 4.7 &  \\
$\mathrm{H_{2}}$ boiler & 175 & 3.5 & \\
\bottomrule
\end{tabular}
\end{center}
\end{table}

\clearpage

\bibliographystyle{bibliography}
\bibliography{SI.bib}